\documentclass[manuscript,natbib=false]{acmart}
\setcopyright{none}
\usepackage[datamodel=acmdatamodel, style=acmnumeric]{biblatex}
\usepackage{tikz}
\usepackage{booktabs}
\usepackage{tabularx}
\usepackage{array}
\usepackage{ctable} 
\usepackage{multirow} 
\usepackage{balance} 
\usepackage{amsmath} 
\usepackage{flushend}
\usetikzlibrary{positioning,arrows.meta}
\usepackage[inline]{enumitem}
\setlist[enumerate]{leftmargin=*}
\usepackage{cleveref}

\AtBeginDocument{%
 }

\usepackage[nolist]{acronym}
\begin{document}

\title{Toward a Unified Security and Privacy Framework for AI-Native 6G Networks}
\author{Bidushi Barua}
\affiliation{%
  \institution{Department of Computer Science, University of York}
  \city{York}
  \country{UK}}
\email{bidushi.barua@york.ac.uk}
\author{Ahsan Khan} 
\affiliation{%
  \institution{Department of Computer Science, University of York}
  \city{York}
  \country{UK}}
\author{Kangfeng Ye}
\affiliation{%
  \institution{Department of Computer Science, University of York}
  \city{York}
  \country{UK}}
\author{Panagiotis Papanastasiou}
\affiliation{%
  \institution{Department of Computer Science, University of York}
  \city{York}
  \country{UK}}
  \author{Yifan Liu}
\affiliation{%
  \institution{Department of Computer Science, University of York}
  \city{York}
  \country{UK}}
\author{Mohit Bidikar}
\affiliation{%
  \institution{Department of Computer Science, University of York}
  \city{York}
  \country{UK}}
\author{Anthony Moulds}
\affiliation{%
  \institution{Department of Computer Science, University of York}
  \city{York}
  \country{UK}}
\author{Julie McCann}
\affiliation{%
  \institution{Department of Computing, Imperial College London}
  \city{York}
  \country{UK}}
\author{Poonam Yadav}
\affiliation{%
  \institution{Department of Computer Science, University of York}
  \city{York}
  \country{UK}}

\renewcommand{\shortauthors}{Barua, et al.}

\begin{abstract}
Sixth Generation (6G) communication networks are expected to evolve into AI-native, highly autonomous ecosystems that integrate communication, computing, sensing, and artificial intelligence. While these capabilities enable unprecedented connectivity and intelligent services, they also create a highly heterogeneous security and privacy landscape that cannot be addressed through isolated, technology-specific solutions. This paper presents a comprehensive survey of security and privacy in AI-native 6G networks from a cross-layer perspective. We first examine the fragmentation of existing security and privacy approaches across emerging technologies, network architectures, AI systems, and standardization efforts, motivating the need for a unified security and privacy framework. Building upon this framework, we develop a cross-layer threat taxonomy encompassing infrastructure, network and architectural, AI, privacy, and security management domains, and analyze representative threats across key AI-native 6G technologies. Furthermore, we map these threats to corresponding cross-layer countermeasures, including standards harmonization as a security function, and identify critical research gaps and future priorities for secure, interoperable, and trustworthy AI-native 6G ecosystems. Finally, we discuss future research directions toward realizing secure, privacy-preserving, resilient, and globally interoperable 6G networks. This survey provides researchers, practitioners, and standardization communities with a holistic foundation for the design, evaluation, and deployment of trustworthy AI-native 6G systems. 
\end{abstract}


%\ccsdesc[300]{Do Not Use This Code~Generate the Correct Terms for Your Paper}
%\ccsdesc{Do Not Use This Code~Generate the Correct Terms for Your Paper}
%\ccsdesc[100]{Do Not Use This Code~Generate the Correct Terms for Your Paper}

\maketitle
\begin{acronym}[CF-mMIMO] % Give the longest label here so that the list is nicely aligned
\acro{3GPP}{3rd Generation Partnership Project}
\acro{5G AKA}{5G Authentication and Key Agreement}
\acro{1G}{1{st} Generation}
\acro{2G}{2{nd} Generation}
\acro{3G}{3{rd} Generation}
\acro{4G}{4{th} Generation}
\acro{5G}{5{th} Generation}
\acro{6G}{6{th} Generation}
\acro{6G-IA}{6G Smart Networks and Services Industry Association}
\acro{AES}{Advanced Encryption Standard}
\acro{AI}{Artificial Intelligence}
\acro{AMF}{Access and Mobility Management Function}
\acro{AML}{Adversarial Machine Learning}
\acro{AN}{Artificial Noise}
\acro{AP}{Access Point}
\acro{APT}{Advanced Persistent Threat}
\acro{ARPF}{Authentication Credential Repository and Processing Function}
\acro{AS}{Access Stratum}
\acro{AUSF}{Authentication Server Function}
\acro{AWGN}{Additive White Gaussian Noise}
\acro{B5G}{Beyond 5G}
\acro{BAN}{Body Area Network}
\acro{BCI}{Brain Computer Interfaces}
\acro{CF-mMIMO}{Cell-free massive MIMO}
\acro{CLID}{Cross Layer Intrusion Detection System}
\acro{COTS}{ Commercial Off-the-Shelf}
\acro{CPRI}{Common Public Radio Interface}
\acro{CSI}{Channel State Information}
\acro{D2D}{Digital-to-Digital}
\acro{DDOS}{Distributed Denial-of-Service}
\acro{DFL}{Decentralized Federated Learning}
\acro{DNN}{Deep Neural Network}
\acro{DoS}{Denial-of-Service}
\acro{DDoS}{Distributed Denial-of-Service}
\acro{DP}{Differential Privacy}
\acro{DL}{Deep Learning}
\acro{DSSS}{Direct-Sequence Spread-Spectrum}
\acro{DT}{Digital Twin}
\acro{EAP}{Extensible Authentication Protocol}
\acro{ECC}{Elliptic-curve cryptography}
\acro{ECU}{Electronic Control Unit}
\acro{eMBB}{ Enhanced Mobile Broadband}
\acro{ETSI}{European Telecommunications Standards Institute}
\acro{FHSS}{Frequency-Hopping Spread-Spectrum}
\acro{FL}{Federated Learning}
\acro{GCOT}{Global Coalition on Telecoms}
\acro{GDPR}{General Data Protection Regulation}
\acro{gNB}{Next Generation Node B}
\acro{HAPS}{High-Altitude Platforms}
\acro{HN}{Home Network}
\acro{6G-IA}{6G Smart Networks and Services Industry Association}
\acro{IDS}{Intrusion Detection System}
\acro{IETF}{Internet Engineering Task Force}
\acro{IMT-2030}{International Mobile Telecommunications 2030} 
\acro{IoE}{Internet of Everything}
\acro{IoST}{Internet of Space Things}
\acro{IoT}{Internet of Things}
\acro{IoV}{Internet of Vehicles}
\acro{ISG}{Industry Specification Group}
\acro{ITU}{International Telecommunication Union}
\acro{ITU-R}{International Telecommunication Union Radiocommunication Sector} 
\acro{ITU-T}{International Telecommunication Union Telecommunication Standardization Sector}
\acro{IMT}{International Mobile Telecommunications}
\acro{JCAS}{Joint Communication and Sensing}
\acro{KAUSF}{Key [from] Authentication Server Function}
\acro{KPI}{Key Performance Indicator}
\acro{LDPC}{Low-Density Parity-Check}
\acro{MEC}{Multi-access Edge Computing}
\acro{MIMO}{Multiple-Input Multiple-Output}
\acro{MITM}{Man-in-the-Middle}
\acro{MISO}{Multiple-Input Single-Output}
\acro{ML}{Machine Learning}
\acro{mMIMO}{Massive Multiple-Input Multiple-Output}
\acro{mMTC}{Massive Machine-Type Communications}
\acro{mmWave}{Millimeter Wave Communication}
\acro{NAS}{Non-Access Stratum}
\acro{NEF}{Network Exposure Function}
\acro{NESA}{National Electronic Security Authority}
\acro{NGA}{North America's Next G Alliance}
\acro{NGMN}{Next Generation Mobile Network Alliance 
}
\acro{NFV}{Network Function Virtualization}
\acro{NIST}{National Institute of Standards and Technology}
\acro{NRF}{Network Repository Function}
\acro{NSA}{Non-Standalone}
\acro{NOMA}{Non Orthogonal Multiple Access}
\acro{NTN}{Non-Terrestrial Networks}
\acro{O-RAN}{Open Radio Access Network}
\acro{OBU}{On-Board Unit}
\acro{P2D}{Physical-to-Digital}
\acro{PII}{Personally Identifiable Information}
\acro{PLKG}{Physical Layer Key Generation}
\acro{PLA}{Physical Layer Authentication}
\acro{PLS}{Physical Layer Security}
\acro{PKI}{Public Key Infrastructure}
\acro{PUF}{Physically Unclonable Function}
\acro{PQC}{Post Quantum Cryptography}
\acro{QKD}{Quantum Key Distribution}
\acro{RAN}{Radio Access Network}
\acro{RF}{Radio Frequency}
\acro{RIC}{RAN Intelligent Controller}
\acro{RIS}{Reconfigurable Intelligent Surfaces}
\acro{RL}{Reinforcement Learning}
\acro{Near-RT}{Near-Real Time}
\acro{Non-RT}{Non-Real Time}
\acro{RSSI}{Received Signal Strength Indicator}
\acro{RSA}{Rivest-Shamir-Adleman}
\acro{RSS}{Received Signal Strength}
\acro{RSU}{Road-Side Unit}
\acro{SA}{Stand-alone}
\acro{SBA}{Service-based Architecture}
\acro{SDN}{Software-Defined Networking}
\acro{SEAF}{Security Anchor Function}
\acro{SEPP}{Security Edge Protection Proxy}
\acro{SIDF}{Subscription Identifier De-concealing Function}
\acro{SINR}{Signal-to-Interference-plus-Noise Ratio}
\acro{SISO}{Single-Input Single-Output}
\acro{SMO}{Service Management and Orchestration}
\acro{SMS}{short message service}%{Short Message Service}
\acro{SN}{Serving Network}
\acro{SNR}{Signal-to-Noise Ratio}
\acro{SPA}{Security Policy Administrator}
\acro{SPCTM}{Sensing Policy, Consent, and Transparency Management Module}
\acro{SPE}{Security Policy Engine}
\acro{SPEP}{Security Policy Enforcement Point} 
\acro{SUCI}{Subscription Concealed Identifier}
\acro{SUPI}{Subscription Permanent Identifier}
\acro{THz}{Terahertz Communication}
\acro{UAV}{Unmanned Aerial Vehicle}
\acro{UDM}{Unified Data Management}
\acro{UE}{User Equipment}
\acro{UICC}{Universal Integrated Circuit Card}
\acro{UPF}{User Plane Function}
\acro{URLLC}{Ultra-Reliable and Low Latency Communication}
\acro{USIM}{Universal Subscriber Identity Module}
\acro{V2I}{Vehicles-to-Infrastructure}
\acro{V2N}{Vehicles-to-Networks}
\acro{V2P}{Vehicles-to-Pedestrians}
\acro{V2V}{Vehicles-to-Vehicles}
\acro{V2X}{Vehicle-to-Everything Communication}
\acro{VANET}{Vehicular Adhoc Network}
\acro{VLC}{Visible Light Communication}
\acro{VM}{Virtual Machines}
\acro{VPN}{Virtual Private Network}
\acro{VPN}{Virtual Private Network}
\acro{WG}{Working Group}
\acro{XR}{extended reality}
\acro{ZTA}{Zero Trust Architecture}
\acro{ZSM}{Zero Touch Network and Service Management}
\end{acronym}

\section{Introduction}\label{sec:intro}
Wireless communication networks have evolved from voice-centric systems to intelligent, highly connected infrastructures supporting broadband services, massive machine-type communications, and ultra-reliable low-latency applications~\cite{nguyen2021security,yang20245g6gsurveysecurity}. Building upon this evolution, \ac{6G} is envisioned as an AI-native network that integrates communication, sensing, computing, and artificial intelligence to enable connected intelligence across diverse applications and services~\cite{saad2019vision}.

However, \ac{5G} is insufficient to support emerging applications requiring ultra-low latency, ultra-high reliability, pervasive intelligence, and real-time decision making. Future services, including \ac{XR}, digital twins, autonomous systems, and the \ac{IoE}, demand seamless integration of communication, sensing, computing, and AI, driving the evolution toward adaptive, intelligent, and autonomous \ac{6G} networks~\cite{yang20245g6gsurveysecurity,nguyen2021security,saad2019vision}.
To support emerging intelligent applications, \ac{6G} is envisioned as an AI-native network that tightly integrates communication, sensing, computing, and intelligence across terrestrial, aerial, satellite, and edge infrastructures. Enabled by technologies such as RIS, JCAS, NTNs, O-RAN, network slicing, edge intelligence, and foundation models, AI-native 6G is expected to deliver substantial improvements in capacity, latency, reliability, and energy efficiency~\cite{HexaXII_D12_2023,ITU-R-M2160}. While this convergence enables unprecedented capabilities, it also fragments the security and privacy landscape, as heterogeneous technologies, AI-driven network functions, and distributed infrastructures introduce interdependent threats that cannot be addressed through isolated security mechanisms.

Unlike traditional threat models that primarily focus on communication networks, AI-native 6G systems are vulnerable to a broader spectrum of attacks targeting physical infrastructures, virtualized network functions, AI models, sensing environments, privacy-sensitive data, and security management mechanisms. Furthermore, the increasing reliance on distributed intelligence, autonomous decision making, and cross-domain interoperability creates complex interdependence among technologies, making security and privacy risks more difficult to identify, isolate, and mitigate. Consequently, threats originating in one layer may propagate across multiple domains, leading to cascading impacts on network reliability, privacy preservation, trust establishment, and service resilience.

Although extensive research efforts have investigated 6G security and privacy (as described in Table \ref{tab:related_work}), existing surveys primarily address individual technologies or specific security mechanisms, leaving a limited understanding of how threats, countermeasures, and standardization efforts interact across AI-native 6G ecosystems. In parallel, standards development organizations (e.g., \ac{ITU}, \ac{3GPP}, \ac{ETSI}, and \ac{IETF}) and industry alliances (e.g., O-RAN Alliance, AI-RAN Alliance, \ac{NGMN}, and Next G Alliance) are developing complementary security and privacy requirements, resulting in an evolving but fragmented landscape.

To address this gap, this survey presents a cross-layer perspective on 6G security and privacy. Specifically, we analyze the fragmentation of current security and privacy approaches, and propose a cross-layer threat taxonomy encompassing infrastructure threats, network and architectural threats, AI and intelligence threats, privacy threats, and security management threats. Building upon this taxonomy, we identify standardization gaps, synthesize emerging countermeasures, and discuss the foundations of a unified security and privacy framework for AI-native 6G networks.

The main contributions of this survey are summarized as follows:
\begin{itemize}
\item We review the evolving vision of AI-native 6G networks and identify the fragmentation challenges associated with security and privacy.
\item We analyze security and privacy standardization efforts, emerging policy initiatives, and harmonization challenges across major standards organizations, industry alliances, and research initiatives.
\item We propose a cross-layer threat taxonomy encompassing infrastructure, network and architectural, AI, privacy, and security management threats.
\item We propose a cross-layer countermeasure framework comprising identity and access management, security orchestration and automation, AI governance and assurance, privacy governance and compliance, and standards harmonization, and identify the associated open research challenges.
\item We present a unified security and privacy framework for AI-native \ac{6G} networks that integrates the proposed taxonomy, and countermeasures.
\end{itemize}
\begin{table*}[ht]
\centering
%\caption{Comparison of Existing Survey Papers on 6G Security} 
\caption{Comparative Analysis of Existing Literature on 6G Security and Privacy}

\label{tab:related_work}
\resizebox{\textwidth}{!}{%
\begin{tabular}{|p{0.6cm}|p{3.8cm}|p{4.2cm}|p{4.5cm}|p{6.8cm}|}
\hline
\textbf{Ref} & \textbf{Focus Area} & \textbf{Enablers Covered} & \textbf{Threat Domains} & \textbf{Limitations} \\
\hline
\cite{yang20245g6gsurveysecurity} & Standardization and 6G security threats & RIS, THz, ML, AI-native infra & General threat types, trust models, DoS, spoofing & Lacks taxonomy, no detailed treatment of application-layer security  \\
\hline
\cite{nguyen2021security} & Prospective 6G technologies and associated risks & RIS, THz, AI/ML, Blockchain & Spoofing, poisoning, eavesdropping, trust models & No application-layer security, lacks taxonomy and integrated defense framework \\
\hline
\cite{ahmad2019security} & Security overview for \ac{5G} and beyond & Network slicing, SDN, NFV, blockchain & Eavesdropping, jamming, rogue slicing, side-channel attacks & 5G-centric; lacks detailed 6G enabler analysis and forward-looking security models \\
\hline
\cite{abdel2022security} & Security in B5G and 6G communication networks & UAVs, THz, RIS, blockchain & Privacy leakage, spoofing, data integrity, authentication threats & Lacks structured taxonomy and defense mapping; generalist perspective \\
\hline
\cite{naeem2023security} & Security and privacy for RIS in 6G & RIS (reflection control, deployment) & Eavesdropping, spoofing, signal manipulation & Narrow focus on RIS; lacks integration with broader 6G architecture or enabler inter-dependencies \\
\hline
\cite{kim2023security} & FL and blockchain-based security in 6G V2X & FL, blockchain, V2X, edge AI & Model poisoning, Sybil attacks, data tampering & Focused on V2X; lacks generalizability across other 6G use cases or layers \\
\hline
\cite{porambage2024security} & Security, privacy, and trust in O-RAN for 6G & RIC, ZTA, AI anomaly detection, blockchain & Rogue xApps, data leakage, interface tampering & Focused on O-RAN; lacks integration with other 6G domains and cross-layer threats \\
\hline
\cite{nguyen2024emerging} & Enabling technologies and challenges in 6G NTNs & Satellites, UAVs, HAPS, RIS & Handover failures, dynamic spectrum risks, latency threats & Lacks focus on E2E security, privacy-preserving methods, and zero-trust integration \\
\hline
\cite{ferrag2023edge} & Edge learning security for 6G-enabled IoT & Federated Learning, Edge AI & Data poisoning, backdoors, adversarial examples, privacy leakage & Focuses only on IoT edge learning; lacks broader enabler and architectural security coverage \\
\hline
\cite{mao2023security} & Security and privacy at 6G network edge & Edge computing, zero-trust, immersive services & Identity spoofing, inference leakage, access control, trust models & Lacks system-wide threat taxonomy and enabler diversity (e.g., RIS, JCAS) \\
\hline
\cite{nguyen2021federated} & FL integration in IoT with privacy/security focus & FL, blockchain, differential privacy & Poisoning, inference, model manipulation & Focused on IoT-FL; lacks integration with broader 6G enablers and cross-layer security view \\
\hline
\cite{yuan2024decentralized} & Decentralized federated learning & DFL, peer collaboration, FL over edge & Poisoning, Sybil attacks, unstable convergence & Strong DFL focus; lacks broader 6G enabler context and system-level integration \\
\hline
\cite{farsimadan2025review} & Security review for V2X in VANETs & VANETs, V2V, V2I, encryption, trust models & Sybil attacks, message falsification, DoS, privacy leakage & Limited to VANETs; lacks alignment with 6G enablers and future-proof models \\
\hline
\cite{9961877} & RL-based physical and cross-layer security in 6G & RIS, THz, RL, edge AI & Jamming, spoofing, eavesdropping & Focused on RL methods; lacks broad threat taxonomy and architectural-level security analysis \\
\hline
\cite{de2023survey} & Role of physical layer security in 6G & PLS, AN injection, beamforming, secure CSI & Eavesdropping, jamming, spoofing & Strong PHY focus; lacks cross-layer integration and AI-driven adaptability \\
\hline
\cite{dass2024addressing} & Security of network slicing in 5G/6G & AI-based slice monitoring, blockchain, secure orchestration & Slice hijacking, isolation failure, rogue slice & Focused on slicing; lacks integration with other 6G enablers and holistic architecture \\
\hline
\cite{10935636} & AI convergence with 6G communication networks & Semantic comm., beamforming, threat detection & AI-based attacks, lack of transparency, adversarial examples & Focused on AI integration; lacks detailed privacy models and threat taxonomy \\
\hline
This work & Unified Security and Privacy Objectives, Cross-Layer Threat Taxonomy and Countermeasures in AI-native 6G & RIS, JCAS, NTN, THz,XL-MIMO, O-RAN, Network Slicing, iZTA, IBN, MEC, FL, AI-RAN, AI agents, LLMs, XR and Metaverse, Semantic Communications, PQC, Secure Boot and Remote Attestation & Infrastructure, Network\&Architecture, AI/Intelligence, Privacy and Security Management Threats and Fragmentation Across Networks &  \\
\hline
\end{tabular}%
}
\end{table*}

\section{Fragmentation in 6G Security and Privacy}
\begin{figure}[t]    
\centering
\includegraphics[width=\columnwidth]{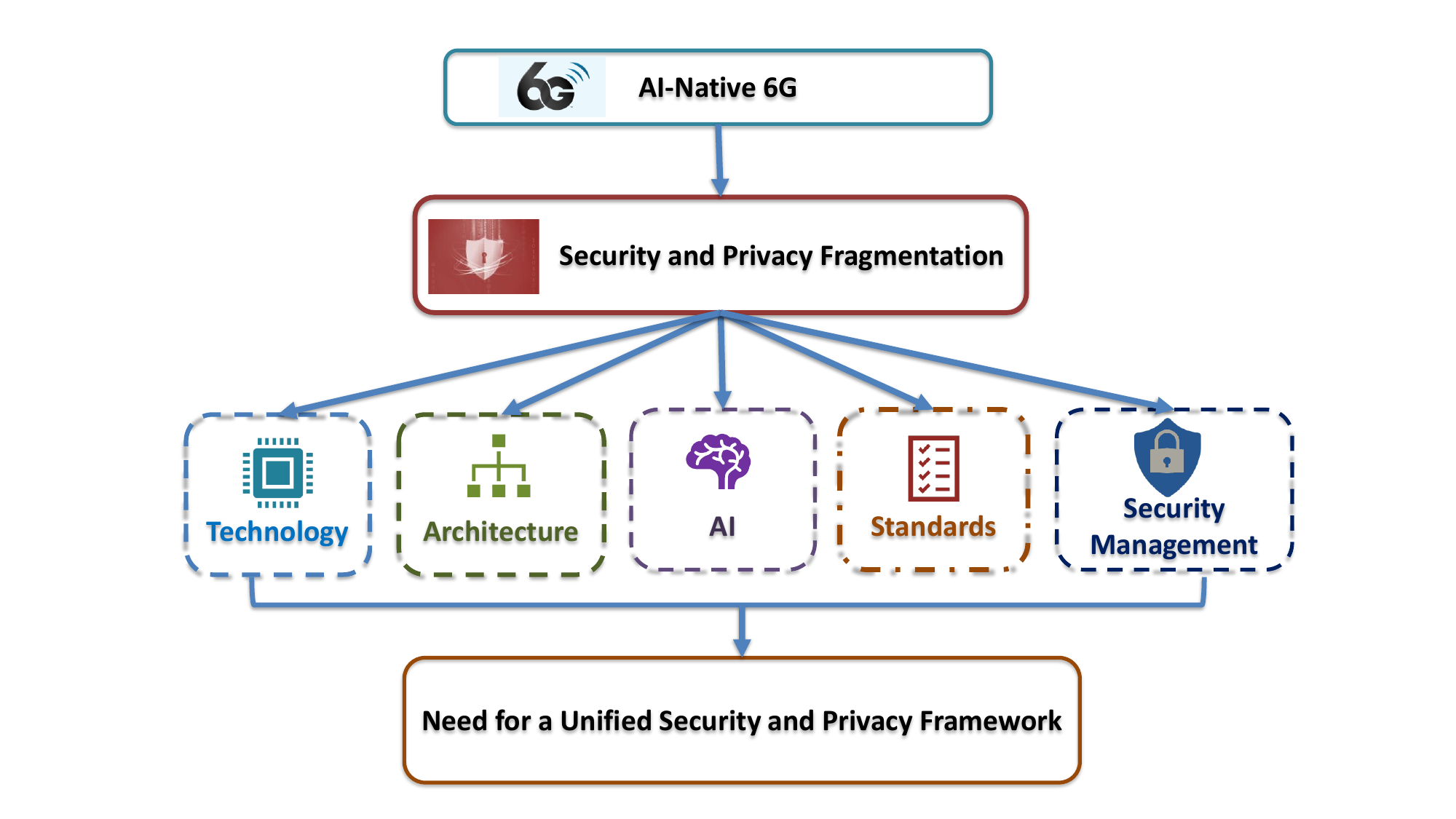}
\caption{Fragmentation in Security and Privacy of AI-native 6G Networks.} \label{fig:fragmentation_framework}
    \end{figure}
AI-native 6G introduces fragmented security and privacy challenges through the convergence of communication, sensing, computing, and AI. Heterogeneous technologies, distributed architectures, AI-driven decision making, and evolving standards create inconsistencies in security assumptions, protection mechanisms, and governance models. As a result, developing interoperable and end-to-end security solutions for AI-native 6G remains a significant challenge.
The major sources of fragmentation can be broadly categorized into five dimensions:

\begin{itemize}
\item \textbf{Technology Fragmentation:} Diverse technologies such as RIS, JCAS, NTN, THz communications, and digital twins introduce heterogeneous attack surfaces.

\item \textbf{Architectural Fragmentation:} O-RAN, network slicing, edge computing, and cloud-native infrastructures create distributed trust boundaries and management complexities.

\item \textbf{AI Fragmentation:} Federated learning, foundation models, AI-RAN, and autonomous agents introduce new threats related to adversarial AI and autonomous decision making.

\item \textbf{Standards Fragmentation:} Multiple standardization bodies define security requirements from different perspectives, leading to interoperability challenges.

\item \textbf{Security Management Fragmentation:} Diverse approaches to identity management, authentication, and quantum-safe security create inconsistencies across domains.
\end{itemize}These fragmentation dimensions collectively contribute to a highly interconnected threat landscape in which vulnerabilities originating in one domain can propagate across multiple layers. Consequently, a cross-layer perspective is required to systematically identify, analyze, and mitigate security and privacy risks in AI-native 6G networks.
\begin{figure}[t]    
\centering
\includegraphics[width=\columnwidth]{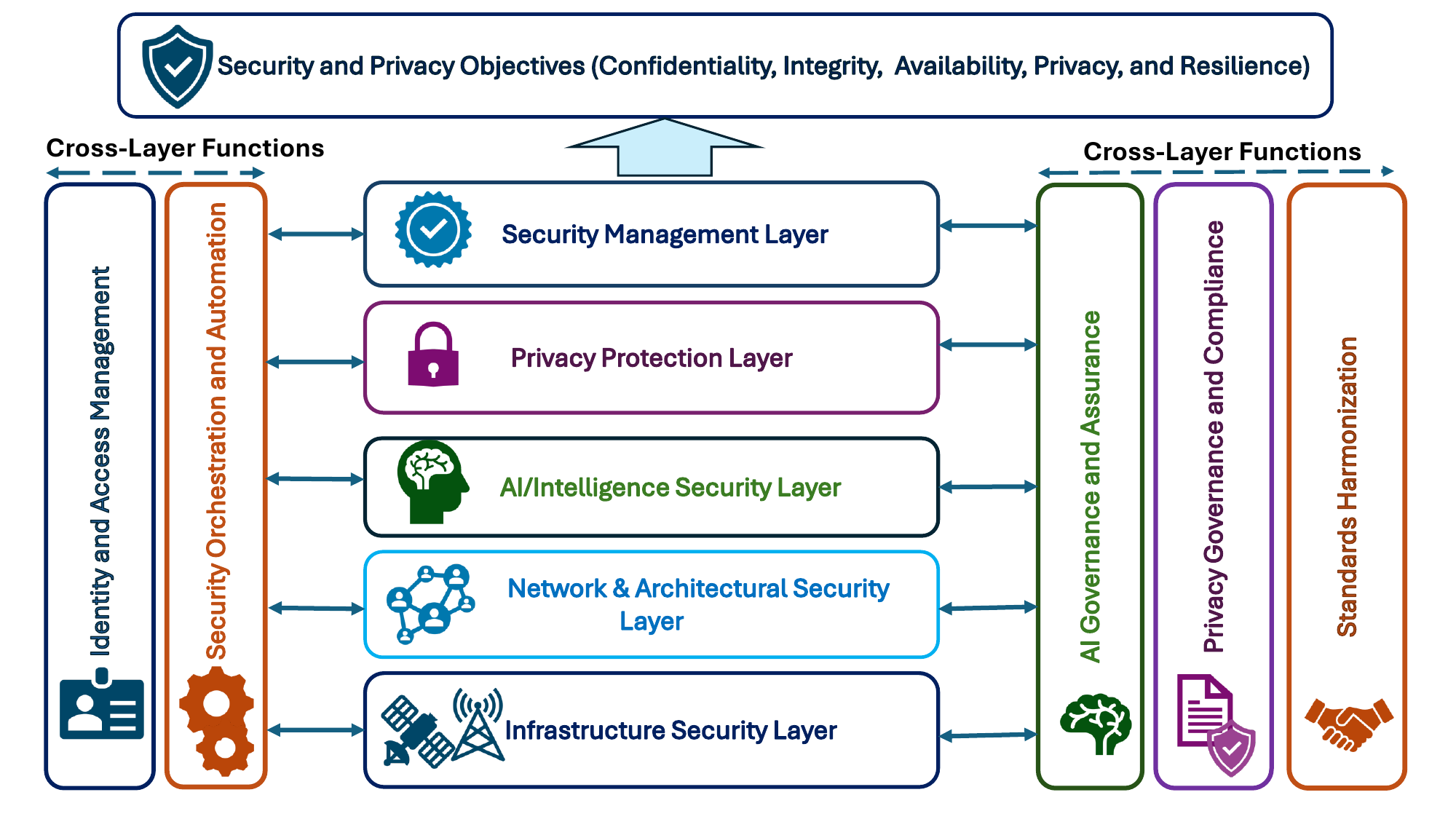}
\caption{Proposed Unified Security and Privacy Framework for AI-Native 6G Networks.} \label{fig:net_sec3}
    \Description{6G architecture}
\end{figure}
Figure~\ref{fig:fragmentation_framework} illustrates the major dimensions of fragmentation in 6G security and privacy. These dimensions span technological, architectural, intelligence-driven, and standards-related perspectives, collectively creating a complex and highly dynamic threat landscape.
The fragmentation of security and privacy requirements across heterogeneous technologies, architectures, AI systems, standards, and management domains highlights the need for a holistic protection strategy. To address this challenge, we propose a Unified Security and Privacy Framework for AI-native 6G networks. 
The framework organizes security and privacy requirements into five security domains and five cross-layer security functions that collectively provide end-to-end protection across the 6G ecosystem. The proposed framework serves as the foundation for the subsequent threat taxonomy, standards harmonization, and countermeasure analysis presented in this paper.
\begin{figure}[t]    
\centering
\includegraphics[width=\columnwidth]{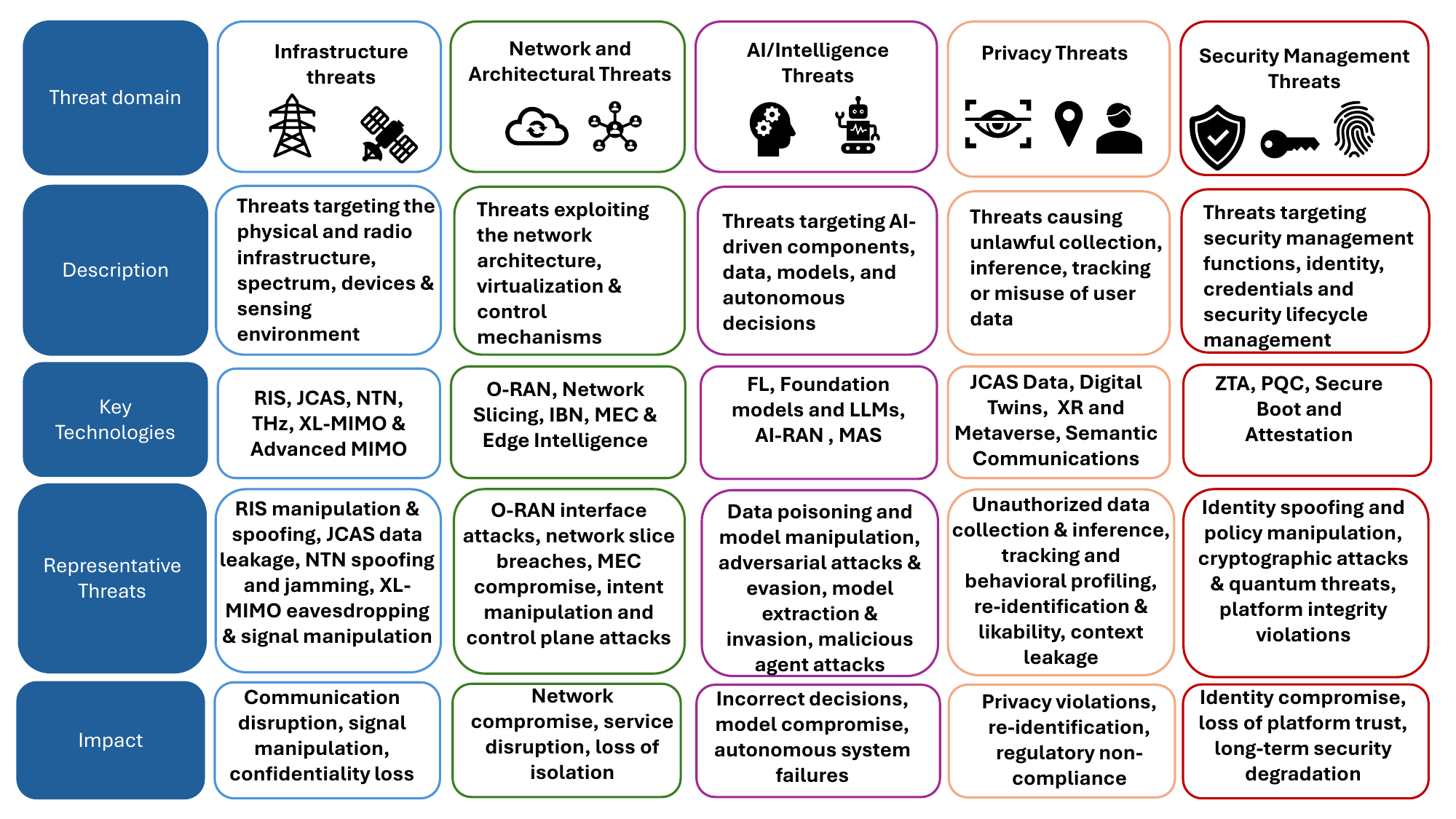}
\caption{Cross-Layer Taxonomy of Threats in AI-Native 6G.} \label{fig:net_sec3}
    \Description{6G architecture}
\end{figure}
\section{Proposed Unified Security and Privacy Objectives for AI-Native 6G Networks}
The proposed Unified Security and Privacy Framework adopts a defense-in-depth approach for AI-native 6G networks by integrating layer-specific security mechanisms with cross-layer security functions. The framework organizes security and privacy requirements into five security domains, namely Infrastructure Security, Network and Architectural Security, AI Security, Privacy Protection, and Security Management, which collectively address the diverse threat landscape identified in the proposed cross-layer taxonomy. To ensure end-to-end protection, five cross-layer functions—Identity and Access Management, Security Orchestration and Automation, AI Governance and Assurance, Privacy Governance and Compliance, and Standards Harmonization—operate across all layers to provide continuous authentication, policy enforcement, risk management, accountability, and regulatory compliance. Through the coordinated interaction of these components, the framework aims to achieve key security and privacy objectives, including confidentiality, integrity, availability, privacy preservation, resilience, trustworthiness, and compliance across heterogeneous AI-native 6G ecosystems.
\section{Cross-Layer Threat Taxonomy}
The heterogeneous and AI-native nature of 6G networks introduces a diverse threat landscape that extends beyond traditional communication systems. Due to integration of advanced communication technologies, distributed intelligence, autonomous decision-making, integrated sensing, and highly decentralized infrastructures, 6G networks create new attack surfaces across multiple layers of the network. To systematically analyze these threats and identify their security and privacy implications, we propose a cross-layer threat taxonomy that classifies threats into five major categories: \textit{Infrastructure Threats}, \textit{Network Threats}, \textit{AI Threats}, \textit{Privacy Threats}, and \textit{Security Management Threats}. This taxonomy provides a unified framework for understanding the evolving 6G threat landscape and serves as the basis for the subsequent analysis of vulnerabilities, standardization gaps, and security requirements.
\subsection{Infrastructure Layer Threats}
The Infrastructure Layer forms the foundation of AI-native 6G networks and encompasses the physical communication, sensing, and connectivity technologies that enable ubiquitous and ultra-reliable services. Key infrastructure enablers include Reconfigurable Intelligent Surfaces (RIS), Joint Communication and Sensing (JCAS), Non-Terrestrial Networks (NTN), and Massive Multiple-Input Multiple-Output (mMIMO) systems. While these technologies significantly enhance network coverage, capacity, intelligence, and sensing capabilities, they also introduce new attack surfaces that extend beyond those encountered in conventional wireless networks. Threats at this layer primarily target the integrity, confidentiality, and availability of physical communications, sensing operations, radio resources, and distributed infrastructure components. The following subsections examine the major security and privacy threats associated with these enabling technologies and discuss their implications for future 6G deployments.
\subsubsection{Reconfigurable Intelligent Surface Threats:}
\begin{table*}[t]
\centering
\caption{Major Infrastructure Threats in RIS-Enabled 6G Networks}
\label{tab:ris_threats}
\renewcommand{\arraystretch}{1.15}
\begin{tabular}{|p{3.5cm}|p{5.5cm}|p{5cm}|}
\hline
\textbf{Threat} &
\textbf{Description} &
\textbf{Potential Impact in 6G Networks} \\
\hline

Passive Eavesdropping
\cite{zhang2021improving}
&
RIS reflections unintentionally expose signals to unauthorized receivers.
&
Disclosure of confidential communications and user information.
\\
\hline

Illegal or Rogue RIS
\cite{khalid2023reconfigurable}
&
Unauthorized RIS devices manipulate signal propagation or redirect traffic.
&
Traffic interception, service disruption, and data leakage.
\\
\hline

Malicious Reconfiguration
\cite{xu2023reconfiguring}
&
Attackers compromise RIS controllers and modify reflection parameters.
&
Beam manipulation, degraded communication quality, and denial of service.
\\
\hline

Control Channel Attacks
\cite{xu2023reconfiguring}
&
Compromise of RIS control signaling and management interfaces.
&
Unauthorized configuration changes and loss of network integrity.
\\
\hline

Side-Channel Information Leakage
\cite{xu2023reconfiguring}
&
Analysis of RIS control signals to infer network behavior.
&
Exposure of operational information and traffic patterns.
\\
\hline

Location and Behavioral Inference
\cite{naeem2023security}
&
RIS-assisted sensing and propagation characteristics reveal user context.
&
Privacy leakage involving location, mobility, and behavioral patterns.
\\
\hline

\end{tabular}
\end{table*}
 Reconfigurable Intelligent Surfaces (RIS) are expected to become key
infrastructure components of AI-native 6G networks, enabling programmable control of wireless propagation environments and improving coverage, reliability, and spectral efficiency. However, their programmable and distributed nature introduces new security and privacy risks beyond those encountered in conventional wireless systems. Adversaries may exploit RIS control mechanisms to manipulate signal propagation, intercept communications, infer user information, or disrupt network operations. Furthermore, the large-scale deployment of AI-orchestrated RIS infrastructures increases the risk of unauthorized reconfiguration and malicious control, highlighting the need for secure RIS management and trust mechanisms. Table~\ref{tab:ris_threats} summarizes the major security and privacy threats associated with RIS-enabled 6G networks.
\subsubsection{Joint Communication and Sensing Security Threats:}\label{sub:JCAS_sec}
\begin{table*}[t]
\centering
\caption{Major Infrastructure Threats in JCAS-Enabled 6G Networks}
\label{tab:jcas_threats}
\renewcommand{\arraystretch}{1.15}
\begin{tabular}{|p{3.5cm}|p{5.5cm}|p{5cm}|}
\hline
\textbf{Threat} &
\textbf{Description} &
\textbf{Potential Impact in 6G Networks} \\
\hline
Sensing Data Manipulation \cite{chorti2022context,osorio2025rise}
&
Attackers tamper with sensing information through spoofing, replay, jamming, or falsified measurements.
&
Incorrect sensing outcomes and compromised decision making.
\\
\hline

Data Poisoning \cite{zhu2023pushing}
&
Malicious sensing data are injected to manipulate sensing models and inference processes.
&
Reduced sensing accuracy and unreliable network intelligence.
\\
\hline

Unauthorized Sensing and Consent Violations \cite{oppo_6Gsecurity,osorio2025rise}
&
Sensing activities are performed without user awareness or explicit consent.
&
Regulatory non-compliance and privacy breaches.
\\
\hline

Authentication and Access Control Attacks \cite{osorio2025rise,gjermundrod2016privacytracker}
&
Unauthorized entities gain access to sensing data or manipulate disclosure mechanisms.
&
Compromised data integrity, confidentiality, and trustworthiness.
\\
\hline

Edge Data Exploitation \cite{dass2024addressing}
&
Sensitive sensing information processed at edge nodes is extracted or misused.
&
Exposure of personally identifiable information and operational data.
\\
\hline

\end{tabular}
\end{table*}
Joint Communication and Sensing (JCAS) is a key enabling
technology for AI-native 6G networks that integrates communication and sensing functionalities within a unified wireless framework. By leveraging shared radio resources, JCAS supports applications such as environmental perception,
localization, autonomous transportation, digital twins, and intelligent infrastructure. However, the simultaneous processing of communication and sensing information introduces unique security and privacy challenges that are not present in conventional wireless systems. Adversaries may exploit sensing capabilities to infer sensitive information, manipulate sensing data, compromise data integrity, or gain unauthorized access to contextual information. Table~\ref{tab:jcas_threats} summarizes the major security and privacy threats associated with JCAS-enabled 6G networks.
\subsubsection{Non-Terrestrial Network Threats:}
\begin{table*}[t]
\centering
\caption{Major Infrastructure Threats in NTN-Enabled 6G Networks}
\label{tab:ntn_threats}
\renewcommand{\arraystretch}{1.15}
\begin{tabular}{|p{3.5cm}|p{5.5cm}|p{5cm}|}
\hline
\textbf{Threat} &
\textbf{Description} &
\textbf{Potential Impact in 6G Networks} \\
\hline

Eavesdropping and Signal Interception
\cite{nguyen2024emerging,10824882}
&
Long-distance broadcast transmissions expose communications to unauthorized interception.
&
Disclosure of sensitive information and loss of communication confidentiality.
\\
\hline

Jamming and Denial-of-Service
\cite{10824882}
&
Adversaries disrupt NTN communication links through interference and signal jamming.
&
Service degradation, communication outages, and reduced network availability.
\\
\hline

Cross-Domain Trust and Authentication Attacks
\cite{abdel2022security}
&
Exploitation of authentication and trust mechanisms across terrestrial, aerial, and satellite domains.
&
Unauthorized access and compromised network interoperability.
\\
\hline

Virtualization and Control Plane Attacks
\cite{nguyen2024emerging,suomalainen2024cybersecurity}
&
Attacks targeting SDN/NFV, O-RAN interfaces, and cloudified space infrastructures.
&
Compromised network control, resource manipulation, and service disruption.
\\
\hline

AI-Induced Threats
\cite{maric2025system,iqbal2023empowering}
&
Poisoning, evasion, model theft, and manipulation of AI-driven NTN operations.
&
Incorrect spectrum allocation, beamforming, and network management decisions.
\\
\hline

Supply Chain and COTS Vulnerabilities
\cite{nguyen2024emerging}
&
Exploitation of vulnerabilities in commercial-off-the-shelf hardware and software components.
&
System compromise and propagation of malicious code across network segments.
\\
\hline

Privacy Leakage
\cite{abdel2022security}
&
Exposure of user, operational, and contextual information across multiple domains.
&
Privacy violations and unauthorized tracking of users and devices.
\\
\hline

\end{tabular}
\end{table*}
 Non-Terrestrial Networks (NTN), comprising satellites, High-Altitude Platform
Stations (HAPSs), and Unmanned Aerial Vehicles (UAVs), are expected to play a crucial role in AI-native 6G networks by
extending connectivity beyond terrestrial infrastructures. Through the integration of space-air-ground networks, NTN
enable global coverage, ubiquitous IoT connectivity, disaster recovery services, and mission-critical communications.
However, their highly distributed architecture, long-distance wireless links, reliance on commercial-off-the-shelf components, virtualization technologies, and increasing use of AI-driven automation introduce significant security and privacy challenges. The heterogeneous and multi-domain nature of NTN further complicates trust establishment, authentication, and end-to-end security. Table~\ref{tab:ntn_threats} summarizes the major security and privacy threats associated with NTN-enabled 6G networks.
\subsubsection{Threats in Emerging Spectrum Technologies:}
\begin{table*}[t] \centering \caption{Major Infrastructure Threats in Emerging Spectrum Technologies for 6G Networks} \label{tab:spectrum_threats} \renewcommand{\arraystretch}{1.15} \begin{tabular}{|p{3.5cm}|p{5.5cm}|p{5cm}|} \hline \textbf{Threat} & \textbf{Description} & \textbf{Potential Impact in 6G Networks} \\ \hline THz Eavesdropping \cite{zhang20196g,osorio2022towards} & Line-of-sight THz transmissions may be intercepted despite highly directional beamforming. & Disclosure of confidential communications and sensitive information. \\ \hline Beam Misalignment and Hijacking \cite{kokkoniemi2020impact} & Adversaries exploit beam training or beam steering procedures to intercept or redirect highly directional transmissions. & Loss of communication confidentiality, degraded link performance, and service disruption. \\ \hline Optical Signal Injection \cite{9482609} & Malicious optical transmitters inject forged VLC/LiFi signals into legitimate communication channels. & False data delivery, communication disruption, and compromised system integrity. \\ \hline Jamming and Interference Attacks \cite{9482609} & Intentional optical or THz interference degrades communication quality and availability. & Reduced throughput, denial of service, and communication outages. \\ \hline Cooperative Eavesdropping \cite{9482609} & Multiple adversaries collaborate to intercept VLC or hybrid RF-optical transmissions. & Enhanced capability to recover confidential information and infer user activities. \\ \hline Hybrid RF-Optical Attack Propagation \cite{khan2026hybrid} & Vulnerabilities in one communication domain (RF or optical) are leveraged to compromise the other. & Cross-domain security breaches and expanded attack surfaces. \\ \hline Privacy Leakage \cite{nguyen2021security} & Communication patterns, localization information, or contextual data reveal user behavior and activities. & Unauthorized profiling, tracking, and privacy violations. \\ \hline \end{tabular} \end{table*}
 Emerging spectrum technologies, including Terahertz (THz) communications and Optical Wireless Communications (OWC) systems such as Visible Light Communications (VLC) and
LiFi, are expected to play a vital role in enabling ultra-high data rates, massive capacity, and low-latency services
in AI-native 6G networks. While these technologies offer highly directional transmissions and improved spectrum
utilization, they also introduce new security and privacy challenges. The unique propagation characteristics of THz and optical channels create vulnerabilities related to eavesdropping, signal interception, jamming, spoofing, and data manipulation. Furthermore, the increasing deployment of hybrid RF-optical systems and public-access communication infrastructures expands the attack surface and complicates security management. Table~\ref{tab:spectrum_threats} summarizes the major security and privacy threats associated with emerging spectrum technologies in 6G networks.

\subsubsection{Extremely Large-Scale MIMO and Advanced MIMO Threats:}
\begin{table*}[t] \centering \caption{Major Infrastructure Threats in XL-MIMO and Advanced MIMO Systems} \label{tab:mmimo_threats} \renewcommand{\arraystretch}{1.15} \begin{tabular}{|p{3.5cm}|p{5.5cm}|p{5cm}|} \hline \textbf{Threat} & \textbf{Description} & \textbf{Potential Impact in 6G Networks} \\ \hline Eavesdropping and Spoofing \cite{mucchi2021physical}& Attackers exploit uplink training and downlink transmissions to intercept communications or impersonate legitimate users. & Compromised confidentiality, unauthorized access, and degraded trust. \\ \hline Pilot Contamination and Injection \cite{irram2022physical}& Malicious manipulation of pilot signals corrupts \ac{CSI} and beamforming processes. & Reduced communication quality and degraded beamforming accuracy. \\ \hline Beamforming Manipulation \cite{cui2022near} & Adversaries influence beam selection or steering mechanisms to redirect or disrupt transmissions. & Service degradation, signal leakage, and denial of service. \\ \hline Adversarial Machine Learning Attacks \cite{catak2022security}& Poisoning, evasion, and model inversion attacks target AI-driven channel estimation and beam prediction systems. & Compromised network optimization and unreliable AI-assisted decisions. \\ \hline Side-Channel Inference \cite{mitev2023physical}& Spatial signal characteristics are exploited to infer user location, mobility patterns, or application behavior. & Privacy leakage and unauthorized user profiling. \\ \hline Jamming and Interference Attacks \cite{yan2016jamming} & Malicious interference targets large-scale antenna arrays and beamforming operations. & Reduced availability and communication reliability. \\ \hline \ac{CSI} Manipulation \cite{cui2022near} & Attackers tamper with channel estimation information used for resource allocation and transmission optimization. & Incorrect scheduling, resource allocation, and communication failures. \\ \hline \end{tabular} \end{table*}
Massive MIMO (mMIMO) technologies are evolving in 6G toward Extremely Large MIMO (XL-MIMO), Cell-Free Massive MIMO, and Distributed MIMO architectures, enabling higher spectral efficiency, enhanced coverage, ultra-precise beamforming, and AI-driven radio resource management. However, the increased spatial resolution, distributed deployment, and integration of AI-driven beam management introduce new security and privacy challenges. Adversaries may target channel estimation procedures, beamforming mechanisms, machine learning models, and wireless propagation characteristics to compromise confidentiality, integrity, and availability. Furthermore, the fine-grained spatial awareness provided by advanced MIMO systems raises concerns regarding user privacy and location inference. Table~\ref{tab:mmimo_threats} summarizes the major security and privacy threats associated with XL-MIMO and advanced MIMO technologies in AI-native 6G networks.

\subsection{Network and Architectural Threats}
The Network and Architectural Layer encompasses the software-defined, cloud-native, and intelligent networking technologies that underpin AI-native 6G systems. 6G networks are expected to rely extensively on open architectures, distributed cloud infrastructures, autonomous network management, and intelligent security mechanisms to support highly dynamic and heterogeneous services. Key architectural enablers include Open Radio Access Networks (O-RAN), Network Slicing, Intent-Based Networking (IBN), Multi-Access Edge Computing (MEC) and Edge Intelligence. While these technologies enhance flexibility, scalability, automation, and service agility, they also introduce new attack surfaces associated with virtualization, orchestration, distributed control, open interfaces, and autonomous decision making. Consequently, vulnerabilities at the network and architectural layer can have system-wide impacts, affecting service availability, network integrity, privacy, and trust. The following subsections examine the major security and privacy threats associated with these architectural enablers and their implications for AI-native 6G deployments.
 \subsubsection{Open RAN Threats:}
\begin{table*}[t]
\centering
\caption{Major Network and Architectural Threats in O-RAN-Enabled 6G Networks}
\label{tab:oran_threats}
\renewcommand{\arraystretch}{1.15}
\begin{tabular}{|p{3.6cm}|p{5.5cm}|p{5cm}|}
\hline
\textbf{Threat} &
\textbf{Description} &
\textbf{Potential Impact in 6G Networks} \\
\hline

Open Interface Attacks \cite{liyanage2023open,groen2024implementing}
&
Exploitation of E2, A1, O1, O2, and Open Fronthaul interfaces through message tampering, unauthorized access, or protocol abuse.
&
Compromised network integrity, information leakage, and service disruption.
\\
\hline

Denial-of-Service (DoS) Attacks \cite{tsourdinis2024ai}
&
Overloading or disrupting open interfaces, RIC functions, or management platforms.
&
Service degradation and reduced network availability.
\\
\hline

Malicious xApps/rApps \cite{porambage2024security}
&
Compromised or malicious third-party applications manipulate RAN control decisions and policies.
&
Incorrect resource allocation, degraded performance, and network instability.
\\
\hline

AI/ML Model Manipulation \cite{porambage2024security,polese2023understanding}
&
Poisoning, evasion, model theft, or adversarial attacks targeting AI-driven control functions within Near-RT and Non-RT RICs.
&
Unreliable network optimization and compromised autonomous decision making.
\\
\hline

Cloud-Native Infrastructure Attacks \cite{shehab2025cloud}
&
Container breakout, privilege escalation, lateral movement, and attacks on virtualized infrastructure.
&
Compromise of hosting platforms and critical network functions.
\\
\hline

Supply Chain Attacks \cite{liyanage2023open,hanselman2020security}
&
Exploitation of vulnerabilities in hardware, software, firmware, or update mechanisms across multiple vendors .
&
Backdoor insertion, unauthorized access, and trust violations.
\\
\hline

API and Orchestration Attacks \cite{polese2023understanding}
&
Abuse of exposed APIs, orchestration platforms, or SMO functions.
&
Unauthorized configuration changes and service manipulation.
\\
\hline

Data Leakage and Privacy Attacks \cite{soltani2022can}
&
Exposure of telemetry, operational, and user-related information across open interfaces.
&
Privacy violations and disclosure of sensitive network information.
\\
\hline

\end{tabular}
\end{table*}
Open RAN (O-RAN) is a key architectural enabler of AI-native 6G networks,
promoting openness, programmability, virtualization, and multi-vendor interoperability through standardized interfaces and disaggregated network functions. While these characteristics enhance flexibility and innovation, they also expand the attack surface compared with traditional vendor-specific RAN deployments. In particular, the integration of cloud-native infrastructures, \ac{SMO} platforms, \ac{Near-RT} and \ac{Non-RT} \acp{RIC}, and AI-driven xApps/rApps introduces new vulnerabilities associated with open interfaces, intelligent control loops, software supply chains, and third-party applications. Consequently, future O-RAN deployments must address threats targeting interface security, AI-enabled network control, virtualization platforms, and security management across heterogeneous ecosystems. Table~\ref{tab:oran_threats} summarizes the major security and privacy threats associated with O-RAN-enabled 6G networks.

\subsubsection{Network Slicing Threats:}
\begin{table*}[t]
\centering
\caption{Major Network and Architectural Threats in Network Slicing for AI-Native 6G Networks}
\label{tab:slicing_threats}
\renewcommand{\arraystretch}{1.15}
\begin{tabular}{|p{3.8cm}|p{5.2cm}|p{5cm}|}
\hline
\textbf{Threat} &
\textbf{Description} &
\textbf{Potential Impact} \\
\hline

Slice Isolation Breaches \cite{de2023survey,foukas2017network}&
Exploitation of vulnerabilities in logical isolation mechanisms between slices. &
Unauthorized access, lateral attacks, and compromise of tenant isolation. \\
\hline

Resource Exhaustion Attacks \cite{de2023survey}&
Malicious consumption of shared compute, storage, or bandwidth resources . &
QoS degradation and denial of service to critical slices. \\
\hline

Orchestration and Management Attacks \cite{escolar2024network, de2023survey}&
Compromise of slice orchestration, life cycle management, or control interfaces . &
Service hijacking, malicious reconfiguration, and network disruption. \\
\hline

Data Leakage and Privacy Attacks \cite{mao2023security,li2025brand}&
Exposure of user, application, or operational data through shared infrastructures . &
Privacy violations and sensitive information disclosure. \\
\hline

Inter-Slice Trust Exploitation \cite{de2023survey}&
Abuse of trust relationships among slices, tenants, or service providers . &
Privilege escalation and cross-slice compromise. \\
\hline

Slice Misconfiguration \cite{de2023survey}&
Incorrect or malicious modification of slice policies and security settings. &
Unauthorized access and weakened security controls. \\
\hline

\end{tabular}
\end{table*}
 Network slicing is a fundamental architectural capability of AI-native 6G networks that
enables multiple virtualized and logically isolated services to coexist on a shared physical infrastructure. While this flexibility supports diverse application requirements and efficient resource utilization, it also introduces new security and privacy challenges arising from virtualization, multi-tenancy, dynamic orchestration, and resource sharing. In particular, vulnerabilities in slice isolation, orchestration platforms, and trust relationships can be exploited to compromise service confidentiality, integrity, and availability. Furthermore, the increasing integration of edge intelligence and AI-enabled services within network slices raises additional concerns regarding data leakage and privacy preservation. Table~\ref{tab:slicing_threats} summarizes the major security and privacy threats associated with network slicing in AI-native 6G networks.

\subsubsection{Intent-Based Networking Threats:}
\begin{table*}[t]
\centering
\caption{Major Network and Architectural Threats in Intent-Based Networking for AI-Native 6G Networks}
\label{tab:ibn_threats}
\renewcommand{\arraystretch}{1.15}
\begin{tabular}{|p{3.5cm}|p{5.5cm}|p{5cm}|}
\hline
\textbf{Threat} &
\textbf{Description} &
\textbf{Potential Impact} \\
\hline

Intent Manipulation \cite{clemm2020network, leivadeas2022survey}&
Malicious modification of user or operator intents before execution.
&
Unauthorized network behavior and policy violations.
\\
\hline

Intent Translation Attacks \cite{clemm2020network}&
Compromise of intent-to-policy translation mechanisms and decision engines.
&
Incorrect network configurations and service disruption.
\\
\hline

Intent Conflict Exploitation \cite{benzaid2020zsm}&
Injection of conflicting or ambiguous intents to trigger unintended actions.
&
Resource misallocation and degraded service performance.
\\
\hline

AI Model Manipulation \cite{wang2025intent}&
Poisoning or evasion attacks targeting AI models used for intent interpretation and orchestration.
&
Compromised autonomous decision making and unreliable automation.
\\
\hline

Orchestration Platform Attacks \cite{leivadeas2022survey}&
Unauthorized access to orchestration and policy management functions .
&
Service hijacking and malicious network reconfiguration.
\\
\hline

Privacy Leakage \cite{clemm2020network}&
Exposure of user intents, operational policies, and contextual information during intent processing.
&
Disclosure of sensitive business and user information.
\\
\hline

Trust and Explainability Failures \cite{benzaid2020zsm}&
Lack of transparency or validation in automated intent execution.
&
Reduced trust in autonomous network operations.
\\
\hline

\end{tabular}
\end{table*}
  Intent-Based Networking (IBN) is emerging as a key architectural enabler of AI-native 6G networks, allowing operators and applications to specify high-level service intents that are automatically translated into network configurations and operational policies. By leveraging AI, automation, and
closed-loop orchestration, IBN enables autonomous network management, dynamic service provisioning, and adaptive
resource optimization. However, the abstraction of network control through intent-driven automation introduces new security and privacy challenges. Adversaries may manipulate intent definitions, compromise intent translation mechanisms, exploit AI-driven decision engines, or target orchestration platforms to influence network behavior. Furthermore, the increasing reliance on AI models and autonomous policy enforcement raises concerns regarding trustworthiness, explainability, and resilience against adversarial manipulation. Table~\ref{tab:ibn_threats} summarizes the major security and privacy threats associated with Intent-Based Networking in AI-native 6G networks.
\subsubsection{Multi-Access Edge Computing and Edge Intelligence Threats:}
\begin{table*}[t]
\centering
\caption{Major Network and Architectural Threats in MEC and Edge Intelligence for AI-Native 6G Networks}
\label{tab:mec_threats}
\renewcommand{\arraystretch}{1.15}
\begin{tabular}{|p{3.5cm}|p{5.5cm}|p{5cm}|}
\hline
\textbf{Threat} &
\textbf{Description} &
\textbf{Potential Impact} \\
\hline

Edge Node Compromise \cite{xiao2019edge}&
Compromise of edge servers hosting applications, network functions, or AI services . &
Service disruption, unauthorized access, and data leakage. \\
\hline

Virtualization and Container Attacks \cite{xiao2019edge, porambage2018survey}&
Exploitation of virtual machines, containers, or orchestration platforms deployed at the edge. &
Privilege escalation and compromise of multiple edge services. \\
\hline

Data Leakage and Privacy Attacks \cite{xiao2019edge}&
Exposure of user, contextual, or operational data processed at edge nodes. &
Privacy violations and disclosure of sensitive information. \\
\hline

AI Model Poisoning and Manipulation \cite{wang2025intent}&
Poisoning or adversarial manipulation of AI models deployed for edge intelligence . &
Incorrect inference and compromised autonomous decisions. \\
\hline

Resource Exhaustion Attacks \cite{xiao2019edge}&
Malicious consumption of computing, storage, or communication resources at the edge. &
QoS degradation and denial of service. \\
\hline

Unauthorized Service Migration \cite{porambage2018survey}&
Manipulation of workload migration and service placement mechanisms . &
Service hijacking and operational disruption. \\
\hline

Trust and Authentication Attacks \cite{cheng2021blockchain}&
Exploitation of weak authentication or trust management among edge nodes and users. &
Unauthorized access and compromised service integrity. \\
\hline

\end{tabular}
\end{table*}
 Multi-Access Edge Computing (MEC) and Edge Intelligence are expected to play a pivotal role in AI-native 6G networks by enabling distributed computing, real-time
analytics, intelligent service orchestration, and low-latency AI applications closer to end users. By hosting network
functions, AI models, digital twins, and data-intensive applications at the network edge, MEC significantly enhances
responsiveness and resource efficiency. However, the distributed and resource-constrained nature of edge environments introduces new security and privacy challenges. Adversaries may target edge nodes, virtualized workloads, AI models, and data processing pipelines to compromise service integrity, confidentiality, and availability. Furthermore, the proximity of edge platforms to end users and IoT devices increases the risk of privacy leakage, unauthorized access, and exploitation of sensitive contextual information. Table~\ref{tab:mec_threats} summarizes the major security and privacy threats associated with MEC and Edge Intelligence in AI-native 6G networks.

\subsection{AI/Intelligence Threats}AI-native 6G networks fundamentally transform network operation by embedding AI into communication, resource management, service orchestration, and autonomous decision making. While AI enables intelligent and adaptive network behavior, it also introduces a new class of security threats targeting AI models, training data, inference processes, and autonomous agents. Unlike conventional cyberattacks, AI-specific threats can manipulate model behavior, compromise decision making, and propagate across distributed AI ecosystems, potentially affecting multiple network functions simultaneously. As AI becomes a core component of AI-native 6G, ensuring the security, robustness, and trustworthiness of intelligent systems throughout the AI life cycle has become a critical research challenge. The following subsections review the major AI and intelligence-related technologies in 6G and analyze their associated security threats.

\subsubsection{Federated Learning Threats:}
\begin{table*}[t]
\centering
\caption{Major AI/Intelligence Threats in Federated Learning for AI-Native 6G Networks}
\label{tab:fl_threats}
\resizebox{\textwidth}{!}{
\begin{tabular}{|p{3.5cm}|p{5.5cm}|p{5cm}|}
\hline
\textbf{Threat} &
\textbf{Description} &
\textbf{Potential Impact in 6G} \\
\hline

Data Poisoning \cite{hallaji2022federated,pmlr-v119-rosenfeld20b}
&
Malicious clients inject manipulated training samples to corrupt the learning process.
&
Reduced model accuracy and unreliable AI-driven network decisions.

\\
\hline

Model Poisoning \cite{liu2024vertical,HALLAJI2023110384}
&
Attackers manipulate gradient updates or model parameters during aggregation.
&
Compromised model integrity and targeted misclassification.
\\
\hline

Backdoor Attacks \cite{9546463}
&
Insertion of trigger patterns that induce malicious model behavior during inference.
&
Hidden vulnerabilities in AI-enabled network services.
\\
\hline

Model Inversion \cite{10024757,10540055}
&
Reconstruction of sensitive training data from shared model updates.
&
Privacy leakage in digital twins, healthcare, and smart-city applications.
\\
\hline

Membership Inference \cite{10429780}
&
Inference of whether a specific data sample participated in training.
&
Disclosure of sensitive user information and participation.
\\
\hline

Gradient Leakage \cite{10429780}
&
Recovery of private information from exchanged gradients.
&
Exposure of user, sensing, and operational data.
\\
\hline

Sybil Attacks \cite{10090432}
&
An adversary creates multiple fake clients to influence model aggregation.
&
Manipulation of global model updates and trust mechanisms.
\\
\hline

\end{tabular}
}
\end{table*}
 Federated Learning (FL) is considered a key enabler of AI-native 6G networks due to its ability to support distributed intelligence while preserving data locality and reducing communication overhead. However, despite keeping raw data on local devices, FL remains vulnerable to a variety of security and privacy attacks. The decentralized nature of FL allows malicious participants to manipulate model training through poisoning attacks, while shared model updates may leak sensitive information through inference and reconstruction attacks. These threats are particularly significant in 6G environments characterized by massive connectivity, heterogeneous edge devices, digital twins, intelligent transportation systems, and mission-critical applications. Table~\ref{tab:fl_threats} summarizes the major security and privacy threats associated with FL in AI-native 6G networks.

\subsubsection{Foundation Model and Large Language Model Threats:}
\begin{table*}[t]
\centering
\caption{Major AI/Intelligence Threats in Foundation Models and LLMs for AI-Native 6G Networks}
\label{tab:llm_threats}
\resizebox{\textwidth}{!}{
\begin{tabular}{|p{3.5cm}|p{5.5cm}|p{5cm}|}
\hline
\textbf{Threat} &
\textbf{Description} &
\textbf{Potential Impact in 6G} \\
\hline

Prompt Injection \&
Jailbreak Attacks \cite{yi2024jailbreak, owasp_llm01_2025} 
&
Malicious prompts manipulate model behavior, override system instructions, or bypass safety mechanisms.
&
Compromised network orchestration, incorrect policy enforcement, disruption of autonomous management.\\
\hline

Training Data Poisoning \&
Backdoor Attacks \cite{zhou2025survey, wang2024badagent} 
&
Adversarial samples inserted during training or fine-tuning create hidden malicious behaviors.
&
Reduced model integrity, unreliable decision making, compromised AI-driven security functions.
\\
\hline

Privacy Leakage \&
Sensitive Information Disclosure \cite{das2025security, li2024llm}
&
Models unintentionally reveal training data, user information, or proprietary knowledge.
&
Exposure of user data, network telemetry, digital twin information, and sensing data.
 \\
\hline

Membership Inference \&
Model Inversion \cite{fu2024membership} 
&
Attackers infer whether specific records were used for training or reconstruct sensitive information.
&
Privacy violations in personalized services, healthcare, ITS, and digital twins.
\\
\hline

Model Extraction \&
Model Theft \cite{zhao2025survey, vassilev2025nistaml} 
&
Repeated queries are used to replicate model functionality or steal proprietary parameters.
&
Loss of intellectual property, replication of AI services, weakened security assurance.
\\
\hline

Retrieval-Augmented Generation (RAG)
Poisoning \cite{owasp_llm01_2025}
&
Manipulation of external knowledge sources used by LLMs.
&
Injection of malicious knowledge into network management and decision support systems.\\
\hline

Excessive Agency \&
Autonomous Decision Manipulation \cite{owasp_llm01_2025, vassilev2025nistaml} 
&
Attackers exploit AI outputs to trigger unintended actions through connected tools or controllers.
&
Compromised resource allocation, slice management, orchestration, and network control.
\\
\hline

Adversarial Prompting \&
Evasion Attacks \cite{goodfellow2015explainingharnessingadversarialexamples, vassilev2025nistaml}
&
Carefully crafted inputs cause incorrect predictions or responses during inference.
&
Bypassing AI-based intrusion detection and security analytics.
 \\
\hline

\end{tabular}
}
\end{table*}
Foundation models and Large Language Models (LLMs) are expected to support autonomous network
management, orchestration, digital twins, and edge intelligence in AI-native 6G networks. However, they introduce novel attack surfaces, including prompt injection and jailbreak attacks, training-data poisoning and backdoors, privacy leakage through inference and extraction attacks \cite{wang2026aisecurity}, model extraction and intellectual property theft, and autonomous decision manipulation. These threats may compromise model integrity, confidentiality, trustworthiness, and network reliability, necessitating robust AI assurance and privacy-preserving safeguards. Table~\ref{tab:llm_threats} summarizes the major security and privacy threats associated with foundation models and LLMs.
\subsubsection{Threats against AI-enabled network operations:}

\begin{table*}[t]
\centering
\caption{Major AI/Intelligence Threats in AI-RAN and Edge Intelligence for AI-Native 6G Networks}
\label{tab:airan_threats}
\renewcommand{\arraystretch}{1.15}
\begin{tabular}{|p{3.5cm}|p{5.5cm}|p{5cm}|}
\hline
\textbf{Threat} & \textbf{Description} & \textbf{Potential Impact in 6G Networks} \\
\hline

Model Poisoning
\cite{de2023survey}
&
Manipulation of AI-RAN training data or model updates.
&
Incorrect AI-driven network decisions.
\\
\hline

Adversarial Evasion
\cite{goodfellow2015explainingharnessingadversarialexamples, vassilev2025nistaml}
&
Crafted inputs cause AI models to generate incorrect predictions or classifications.
&
Bypassing AI-enabled intrusion detection and anomaly detection systems.
\\
\hline

Edge Node Compromise
\cite{feng2026agentic}
&
Compromise of edge servers hosting AI models and network functions.
&
Service disruption, unauthorized access, and data leakage.
\\
\hline

Model Theft and Extraction
\cite{feng2026agentic}
&
Reconstruction or replication of deployed AI models through repeated queries.
&
Intellectual property loss and unauthorized duplication of AI services.
\\
\hline

Privacy Leakage
\cite{wang2026aisecurity}
&
Exposure of user, telemetry, or operational data processed at edge nodes.
&
Disclosure of sensitive information and privacy violations.
\\
\hline

Data Poisoning
\cite{porambage2021roadmap}
&
Manipulation of telemetry or training datasets used by AI models.
&
Reduced model accuracy and unreliable network decisions.
\\
\hline

Resource Exhaustion
\cite{kholmatov2026toward}
&
Consumption of computational or communication resources at edge nodes.
&
Service degradation and denial of AI-assisted network functions.
\\
\hline

AI Decision Manipulation
\cite{owasp_llm01_2025, vassilev2025nistaml}
&
Exploitation of autonomous AI-driven control loops and orchestration systems.
&
Incorrect slice management, resource allocation, and service orchestration.
\\
\hline

\end{tabular}
\end{table*}
 AI-RAN and Edge Intelligence are emerging as fundamental
components of AI-native 6G networks, enabling intelligent radio resource management, autonomous network optimization, predictive maintenance, energy-efficient operation, and real-time service orchestration. By integrating AI directly into the \ac{RAN} and edge infrastructure, these technologies facilitate low-latency
decision-making and distributed intelligence across heterogeneous network environments. However, their extensive    reliance on AI models, edge computing resources, and open network architectures  introduces new security and privacy challenges. Adversaries may target AI training pipelines, edge infrastructure, model repositories, or decision-making processes to manipulate network operations, degrade service quality, or compromise sensitive information. Table~\ref{tab:airan_threats} summarizes the major security and privacy threats associated with AI-RAN and Edge Intelligence in AI-native 6G networks.

\subsubsection{Multi-Agent System Threats:} 
\begin{table*}[t]
\centering
\caption{Major AI/Intelligence Threats in Multi-Agent Systems for AI-Native 6G Networks}
\label{tab:mas_threats}
\renewcommand{\arraystretch}{1.15}
\begin{tabular}{|p{3.5cm}|p{5.5cm}|p{5cm}|}
\hline
\textbf{Threat} &
\textbf{Description} &
\textbf{Potential Impact in 6G Networks} \\
\hline

Agent Impersonation \cite{kholmatov2026toward}
&
Malicious entities masquerade as legitimate agents to gain unauthorized participation.
&
Unauthorized access, trust violations, and manipulation of collaborative decisions.
\\
\hline

Trust Poisoning \cite{nguyen2026security}
&
Attackers manipulate reputation or trust scores used among agents.
&
Selection of malicious agents and degradation of network reliability.
\\
\hline

Malicious Agent Insertion \cite{zheng2025demonstrations}
&
Compromised or rogue agents join the system and participate in coordination tasks.
&
Disruption of orchestration, resource allocation, and service management.
\\
\hline

Collusion Attacks \cite{feng2026agentic}
&
Multiple malicious agents cooperate to influence collective decision making.
&
Biased outcomes and coordinated manipulation of network behavior.
\\
\hline

Communication Manipulation \cite{feng2026agentic}
&
Interception, modification, or spoofing of inter-agent messages.
&
Loss of integrity and reliability of distributed coordination.
\\
\hline

Privacy Leakage \cite{such2014survey}
&
Sensitive information exchanged among agents is exposed or inferred.
&
Disclosure of user, operational, and contextual information.
\\
\hline

Reward Manipulation \cite{wu2023reward}
&
Attackers manipulate feedback or reward signals used for learning.
&
Suboptimal or adversarial agent behavior.
\\
\hline

Autonomous Decision Manipulation \cite{jin2025comprehensive}
&
Exploitation of autonomous planning and decision-making mechanisms.
&
Incorrect orchestration, resource allocation, and service delivery.
\\
\hline

\end{tabular}
\end{table*}
Multi-Agent Systems (MAS) are expected to play a central role in AI-native 6G networks by enabling autonomous coordination among distributed intelligent entities, including network controllers, edge nodes, digital twins, autonomous vehicles, and service orchestrators. Through collaborative decision-making and
task allocation, MAS can support self-organizing networks, intelligent resource management, and adaptive service delivery. However, the distributed and autonomous nature of MAS introduces unique security and privacy challenges. Adversaries may exploit communication channels, manipulate trust relationships, compromise individual agents, or influence collaborative decision-making processes to disrupt network operations. As autonomous agents increasingly interact with critical network functions, ensuring trustworthy coordination and resilient cooperation becomes essential. Table~\ref{tab:mas_threats} summarizes the major security and privacy threats associated with Multi-Agent Systems in AI-native 6G networks.

\subsection{Privacy Threats}
The \ac{6G} applications have very demanding performance requirements, which are provided by highly complex applications with highly malicious actors in the network. This leads to stringent security requirements in these networks. To understand the security and privacy threats in these applications, we discuss the threat landscape for digital twins, and AI agents. Although these applications are not enabling technologies, it is necessary to identify and find countermeasures for making these applications secure and resilient, which finally makes them enablers of secured \ac{6G} technology.
\subsubsection{JCAS Privacy Threats:}
\begin{table*}[t]
\centering
\caption{Major Privacy Threats in JCAS-Enabled AI-Native 6G Networks}
\label{tab:jcas_privacy_threats}
\renewcommand{\arraystretch}{1.15}
\begin{tabular}{|p{3.5cm}|p{5.5cm}|p{5cm}|}
\hline
\textbf{Threat} &
\textbf{Description} &
\textbf{Potential Impact} \\
\hline

Location Tracking \cite{qu2024privacy,lu2024integrated}&
Exploitation of sensing signals to continuously monitor user positions and mobility patterns. &
Loss of location privacy and unauthorized surveillance. \\
\hline

Behavioral Profiling \cite{aakesson2024privacy,dass2024addressing,qu2024privacy}&
Inference of user activities, habits, and routines from sensing observations. &
User profiling and privacy erosion. \\
\hline
Biometric Information Leakage \cite{oppo_6Gsecurity,qu2024privacy}&
Extraction of physiological attributes such as heart rate, gait, or motion signatures from sensing data. &
Disclosure of sensitive personal and health-related information. \\
\hline

Unauthorized Sensing \cite{dass2024addressing,osorio2025rise}&
Collection of environmental or personal information without user knowledge or consent. &
Violation of privacy regulations and user autonomy. \\
\hline

Identity Re-identification \cite{aakesson2024privacy,dass2024addressing}&
Correlation of sensing data with external information sources to identify individuals. &
Loss of anonymity and personal privacy. \\
\hline

Information Leakage and Eavesdropping \cite{lu2024integrated,aakesson2024privacy,gunlu2022secure,oppo_6Gsecurity}
&
Communication and sensing operations expose sensitive channel, localization, or contextual information.
&
Disclosure of user location, mobility patterns, and confidential communications.
\\
\hline
\end{tabular}
\end{table*}

As discussed in subsection \ref{sub:JCAS_sec}, JCAS integrates sensing and communication functionalities within a unified wireless framework. However, the pervasive and often passive nature of sensing introduces significant privacy concerns. Unlike conventional communication systems, JCAS can continuously collect, infer, and process contextual information about users and their surroundings, potentially without explicit user awareness. Consequently, adversaries may exploit sensing capabilities to infer sensitive information, track user behavior, reconstruct personal attributes, or conduct large-scale surveillance. Table~\ref{tab:jcas_privacy_threats} summarizes the major privacy threats associated with JCAS-enabled AI-native 6G networks.
\subsubsection{Digital Twin Privacy Threats:}
\begin{table*}[t]
\centering
\caption{Major Privacy Threats in Digital Twins for AI-Native 6G Networks}
\label{tab:dt_privacy_threats}
\renewcommand{\arraystretch}{1.15}
\begin{tabular}{|p{3.5cm}|p{5.5cm}|p{5cm}|}
\hline
\textbf{Threat} &
\textbf{Description} &
\textbf{Potential Impact} \\
\hline

Identity Re-identification \cite{khan2022digital,10974949}&
Correlation of DT data with external information sources to identify users. &
Loss of anonymity and personal privacy. \\
\hline

Behavioral Profiling \cite{nguyen2021digital,khan2022digital}&
Inference of user habits, activities, and preferences from DT observations. &
Unauthorized profiling and surveillance. \\
\hline

Location and Mobility Tracking \cite{10974949,10979998}&
Continuous synchronization enables monitoring of user movements and trajectories. &
Exposure of location privacy and movement patterns. \\
\hline

Sensitive Data Leakage \cite{khan2022digital,wu2021digital}&
Disclosure of personal, operational, or contextual information stored within DTs. &
Privacy violations and information disclosure. \\
\hline

Cross-Domain Data Aggregation \cite{10974949,nguyen2021digital}&
Integration of data from multiple sources enables comprehensive user inference. &
Excessive collection and misuse of personal information. \\
\hline

Inference and Reconstruction Attacks \cite{10979998,wu2021digital} &
Attackers infer hidden attributes or reconstruct sensitive information from DT data streams. &
Exposure of confidential personal and organizational data. \\
\hline

Digital Twin Data Misuse \cite{khan2022digital}&
Unauthorized use or sharing of DT information by third parties. &
Loss of data sovereignty and regulatory compliance risks. \\
\hline

\end{tabular}
\end{table*}
 Digital Twins (DTs) are expected to become a fundamental component of AI
native 6G networks by creating real-time virtual representations of physical entities, users, devices, networks, and
environments. Through continuous synchronization between physical and digital spaces, DTs enable predictive analytics, intelligent automation, network optimization, and immersive digital experiences. However, the extensive collection, aggregation, and processing of contextual information across physical and virtual domains in Digital Twins (DTs) introduce significant privacy concerns. Since DTs maintain highly detailed representations of users and environments, adversaries may exploit these systems to infer sensitive information, reconstruct user identities, track behaviors, or gain unauthorized access to personal and operational data. The integration of AI, edge intelligence, and large-scale sensing further amplifies these risks by enabling sophisticated inference and profiling attacks. Table~\ref{tab:dt_privacy_threats} summarizes the major privacy threats associated with Digital Twins in AI-native 6G networks.
\subsubsection{XR and Metaverse Privacy Threats:}
\begin{table*}[t]
\centering
\caption{Major Privacy Threats in XR and Metaverse Applications for AI-Native 6G Networks}
\label{tab:xr_privacy_threats}
\renewcommand{\arraystretch}{1.15}
\begin{tabular}{|p{3.5cm}|p{5.5cm}|p{5cm}|}
\hline
\textbf{Threat} &
\textbf{Description} &
\textbf{Potential Impact} \\
\hline

Location and Mobility Tracking \cite{wang2022survey, tukur2023metaverse}&
Continuous monitoring of user position, movement patterns, and spatial interactions. &
Loss of location privacy and user tracking. \\
\hline

Biometric Information Leakage \cite{wang2022survey, ruiu2024metaverse}&
Exposure of eye-tracking, facial expressions, gestures, voice, and physiological signals. &
Identity disclosure and biometric profiling. \\
\hline

Behavioral Profiling \cite{tukur2023metaverse, wang2022survey}&
Inference of user preferences, habits, emotions, and activities from XR interactions. &
Unauthorized profiling and targeted manipulation. \\
\hline

Identity Linkage and Re-identification \cite{wang2022survey}&
Correlation of virtual identities with real-world identities using behavioral and contextual data. &
Loss of anonymity and privacy violations. \\
\hline

Digital Twin Data Leakage \cite{wang2022survey, tukur2023metaverse}&
Exposure of personal or environmental information stored in digital twin representations. &
Disclosure of sensitive personal and contextual information. \\
\hline

Contextual Information Inference \cite{wang2022survey}&
Extraction of environmental, social, or situational information from immersive interactions. &
Unauthorized knowledge of user activities and surroundings. \\
\hline

Cross-Platform Data Aggregation \cite{wang2022survey}&
Collection and fusion of user data across multiple XR, AI, and Metaverse services. &
Comprehensive user surveillance and privacy erosion. \\
\hline

\end{tabular}
\end{table*}
Extended Reality (XR) and Metaverse applications are expected to become
prominent AI-native 6G services, enabling immersive human-machine interactions, digital collaboration, virtual environments, and real-time digital experiences. These applications rely on  continuous collection and processing of multi-modal data, including user location, motion trajectories, eye movements, facial expressions, biometric signals, voice interactions, and environmental context. While such data are essential for delivering immersive and personalized experiences, they also introduce significant privacy risks. The integration of AI-driven analytics, edge intelligence, digital twins, and pervasive sensing further increases the possibility of user profiling, identity inference, behavioral tracking, and unauthorized disclosure of sensitive information. Consequently, preserving privacy in XR and Metaverse ecosystems represents a critical challenge for future AI-native 6G networks. Table~\ref{tab:xr_privacy_threats} summarizes the major privacy threats associated with XR and Metaverse applications.
\subsubsection{Semantic Communication Privacy Threats:}
\begin{table*}[t]
\centering
\caption{Major Privacy Threats in Semantic Communications for AI-Native 6G Networks}
\label{tab:semcom_privacy_threats}
\renewcommand{\arraystretch}{1.15}
\begin{tabular}{|p{3.5cm}|p{5.5cm}|p{5cm}|}
\hline
\textbf{Threat} &
\textbf{Description} &
\textbf{Potential Impact} \\
\hline

Semantic Information Leakage \cite{shen2023secure, won2024resource}&
Sensitive information is inferred from transmitted semantic representations rather than raw data. &
Disclosure of private user intentions, preferences, and contextual information. \\
\hline

Context Inference Attacks \cite{guo2024survey, shen2023secure}&
Adversaries exploit semantic context and knowledge graphs to infer user activities and behaviors. &
Unauthorized profiling and behavioral tracking. \\
\hline

Semantic Reconstruction Attacks \cite{wang2024privacy}&
Attackers reconstruct original content from semantic embeddings or latent representations. &
Exposure of sensitive personal and operational information. \\
\hline

Knowledge Base Poisoning \cite{shen2023secure}&
Manipulation of shared semantic knowledge repositories used for semantic reasoning and communication. &
Incorrect inference and privacy violations. \\
\hline

Cross-Domain Correlation Attacks \cite{guo2024survey}&
Semantic information from multiple domains is correlated to identify users or reveal hidden relationships. &
Identity linkage and re-identification risks. \\
\hline

Metadata and Intent Leakage \cite{shen2023secure, guo2024survey}&
Semantic transmission reveals user goals, interests, and communication intent even when payload data remain protected. &
Loss of user privacy and sensitive information disclosure. \\
\hline

LLM-Assisted Semantic Inference \cite{wang2026aisecurity}&
Foundation models exploit semantic content to infer private attributes beyond the transmitted information. &
Large-scale privacy leakage and profiling. \\
\hline

\end{tabular}
\end{table*}
 Semantic Communications (SemCom) are emerging as a key enabling
technology for AI-native 6G networks, aiming to transmit the semantic meaning of information rather than raw data to improve communication efficiency, intelligence, and resource utilization. By incorporating AI
and knowledge-driven processing into communication systems, SemCom enables context-aware services, intelligent
edge applications, and human-centric communications. However, the extraction, representation, and transmission of semantic information introduce unique privacy challenges that extend beyond conventional data confidentiality concerns. Adversaries may infer sensitive contextual information, reconstruct user intentions, exploit semantic knowledge bases, or correlate semantic metadata across multiple sources to reveal private information. Furthermore, the integration of foundation models, edge intelligence, and distributed AI systems into semantic communication frameworks amplifies the risks of privacy leakage and unauthorized inference. Table~\ref{tab:semcom_privacy_threats} summarizes the major privacy threats associated with Semantic Communications in AI-native 6G networks.

\subsection{Security Management and Identity Threats}
Security management is a fundamental pillar of AI-native 6G networks, enabling secure interactions among users, devices, AI agents, edge platforms, network functions, and service providers across highly heterogeneous and decentralized environments. Unlike previous generations, 6G networks are expected to operate through autonomous decision-making, distributed intelligence, cross-domain service orchestration, and pervasive connectivity, making traditional security mechanisms insufficient. Consequently, future networks will increasingly rely on advanced technologies such as Intelligent Zero Trust Architectures (ZTA), Post-Quantum Cryptography (PQC), and Secure Attestation mechanisms to establish, evaluate, and maintain trust throughout the network life cycle. However, these technologies also introduce new attack surfaces, including identity spoofing, credential compromise, quantum-era cryptographic threats, and attacks on trusted execution environments. The following subsections examine the major security and privacy threats associated with these security management technologies in AI-native 6G networks.
\subsubsection{Intelligent Zero Trust Architecture (iZTA) Security Management Threats:}
\begin{table*}[t]
\centering
\caption{Major Security Management Threats in Intelligent Zero Trust Architectures for AI-Native 6G Networks}
\label{tab:izta_threats}
\renewcommand{\arraystretch}{1.15}
\begin{tabular}{|p{3.5cm}|p{5.5cm}|p{5cm}|}
\hline
\textbf{Threat} &
\textbf{Description} &
\textbf{Potential Impact} \\
\hline

Identity Spoofing \cite{chen2023zero,nahar2024survey}&
Impersonation of legitimate users, devices, or AI agents. &
Unauthorized access and security breaches. \\
\hline

Access Control Manipulation \cite{sedjelmaci2023enabling}&
Abuse or modification of adaptive authorization policies. &
Privilege escalation and policy violations. \\
\hline

Cross-Domain Authorization Attacks \cite{ghoraishi5itrust6g}&
Exploitation of trust relationships across multiple administrative domains. &
Unauthorized resource access and service compromise. \\
\hline

Adversarial AI Attacks \cite{wei2023zero,zegarra2023attentive}&
Poisoning or evasion attacks targeting AI-assisted security decisions. &
Incorrect threat detection and access decisions. \\
\hline

Behavioral Profile Manipulation \cite{wei2023zero}&
Falsification of contextual or behavioral attributes used for continuous verification. &
Compromised risk assessment and trust evaluation. \\
\hline

Privacy Leakage \cite{nahar2024survey,garzon2022decentralized}&
Exposure of identity and behavioral information collected for security monitoring. &
Disclosure of sensitive user and operational data. \\
\hline

Credential and Key Compromise \cite{chen2023zero,okika2025assessing}&
Theft or compromise of authentication credentials and cryptographic keys. &
Loss of authentication integrity and secure communications. \\
\hline

\end{tabular}
\end{table*}
 Intelligent Zero Trust Architecture
(iZTA) is emerging as a fundamental security management paradigm for AI-native 6G networks, where billions of
users, devices, AI agents, network functions, and services interact across highly heterogeneous and decentralized
environments. Unlike traditional perimeter-based security approaches, iZTA adopts the principle of “never trust,
always verify,” enabling continuous authentication, authorization, and adaptive access control based on real-time risk assessment and behavioral analytics \cite{chen2023zero,nahar2024survey}. The integration of AI into ZTA further enables dynamic
policy enforcement, automated threat detection, and context-aware security orchestration across terrestrial, aerial,
satellite, and edge domains. However, the increasing reliance on AI-driven decision making, decentralized identity systems, and cross-domain access management introduces new security challenges. Adversaries may exploit identity mechanisms, manipulate access control policies, compromise trust evaluation processes, or target AI-assisted security functions to bypass security controls and gain unauthorized access. Table~\ref{tab:izta_threats} summarizes the major security management threats associated with iZTA in AI-native 6G networks.

\subsubsection{Post-Quantum Cryptography Threats:}
\begin{table*}[t]
\centering
\caption{Major Security Management Threats in Post-Quantum Cryptography for AI-Native 6G Networks}
\label{tab:pqc_threats}
\renewcommand{\arraystretch}{1.15}
\begin{tabular}{|p{3.5cm}|p{5.5cm}|p{5cm}|}
\hline
\textbf{Threat} &
\textbf{Description} &
\textbf{Potential Impact} \\
\hline

Harvest-Now, Decrypt-Later \cite{mosca2015cybersecurity}&
Adversaries store encrypted traffic for future decryption using quantum computers. &
Loss of long-term confidentiality and sensitive information exposure. \\
\hline

Quantum Cryptanalytic Attacks \cite{shor1997polynomial, preskill2018quantum}&
Quantum algorithms such as Shor's algorithm threaten RSA- and ECC-based cryptography. &
Compromise of authentication, key exchange, and digital signatures. \\
\hline

Implementation Attacks \cite{alagic2020status, cherkaoui2024exploring}&
Side-channel, fault-injection, or timing attacks targeting PQC implementations. &
Secret key leakage and cryptographic compromise. \\
\hline

Migration and Interoperability Risks \cite{nist2023migration, enisa2021pqc}&
Security weaknesses arising during migration from classical to post-quantum cryptography. &
Service disruption and configuration vulnerabilities. \\
\hline

Key Management Attacks \cite{alagic2022status}&
Compromise of cryptographic keys and credential management systems supporting PQC deployments. &
Unauthorized access and loss of communication security. \\
\hline

Crypto-Agility Failures \cite{nist2023migration}&
Inability to rapidly replace or update cryptographic algorithms in evolving threat environments. &
Long-term exposure to emerging cryptographic vulnerabilities. \\
\hline

Hybrid Architecture Exploitation \cite{fedorov2023deploying, zeng2024practical}&
Attacks targeting interactions between classical and post-quantum cryptographic mechanisms. &
Reduction of overall system security and trustworthiness. \\
\hline

\end{tabular}
\end{table*}

Post-Quantum Cryptography (PQC) is emerging as a critical security management technology for AI-native 6G networks due to the growing threat posed by quantum computers to conventional public-key cryptographic schemes. Future 6G ecosystems will support long-lived services involving autonomous systems, digital twins, critical infrastructures, and AI-driven applications, where the confidentiality and integrity of data must be preserved over extended periods. In this context, adversaries may exploit ``harvest-now, decrypt-later'' strategies by collecting encrypted communications today and decrypting them once large-scale quantum computers become available \cite{mosca2015cybersecurity,shor1997polynomial}. To address these risks, NIST-standardized post-quantum algorithms such as Kyber and Dilithium are being integrated into next-generation communication systems \cite{alagic2022status}. However, the transition to PQC introduces new security challenges related to cryptographic migration, implementation vulnerabilities, key management, interoperability, and crypto-agility. Table~\ref{tab:pqc_threats} summarizes the major security management threats associated with PQC deployment in AI-native 6G networks.

\subsubsection{Secure Boot and Remote Attestation Threats:}
\begin{table*}[t]
\centering
\caption{Major Security Management Threats in Secure Boot and Remote Attestation for AI-Native 6G Networks}
\label{tab:attestation_threats}
\renewcommand{\arraystretch}{1.15}
\begin{tabular}{|p{3.5cm}|p{5.5cm}|p{5cm}|}
\hline
\textbf{Threat} &
\textbf{Description} &
\textbf{Potential Impact} \\
\hline

Firmware Tampering \cite{schulz2017boot}&
Modification of firmware or bootloader components before system initialization. &
Execution of malicious code and platform compromise. \\
\hline

Attestation Forgery \cite{dave2020sracare}&
Generation of false attestation reports to impersonate trusted platforms. &
Unauthorized access and trust violations. \\
\hline

Trusted Execution Environment Attacks \cite{menetrey2022attestation}&
Exploitation of vulnerabilities in TEEs and hardware security modules. &
Compromise of protected data and execution environments. \\
\hline

Supply Chain Compromise \cite{tan2025advanced}&
Insertion of malicious hardware, firmware, or software during manufacturing or deployment. &
Persistent backdoors and system-wide compromise. \\
\hline

Virtualization and Container Attacks \cite{sierra2020security}&
Manipulation of virtual machines, containers, or network functions hosted on attested platforms. &
Service disruption and integrity violations. \\
\hline

AI Model Integrity Attacks \cite{xing2026towards}&
Modification of AI models or inference pipelines after deployment. &
Incorrect autonomous decisions and degraded trustworthiness. \\
\hline

Credential and Key Compromise \cite{lebedev2018secure}&
Theft of attestation keys, certificates, or platform credentials. &
Loss of trust and authentication failures. \\
\hline

\end{tabular}
\end{table*}
Secure Boot and Remote Attestation are emerging as critical
security management mechanisms for AI-native 6G networks, enabling the verification of platform integrity across
distributed edge, cloud, and network infrastructures. Secure Boot ensures that devices and network functions execute
only authenticated and trusted software during startup, while Remote Attestation allows external entities to verify the
integrity and configuration of hardware, firmware, operating systems, virtualized network functions, and AI services \cite{schulz2017boot, dave2020sracare}. In AI-native 6G environments, these mechanisms are particularly important for securing edge intelligence platforms, O-RAN components, digital twins, autonomous agents, and cloud-native network functions. However, the increasing scale, virtualization, and decentralization of future networks introduce new attack surfaces targeting attestation protocols, trusted execution environments, firmware integrity, and supply-chain trust. Consequently, compromising platform integrity can enable attackers to bypass security controls, manipulate AI services, and gain persistent access to critical network resources. Table~\ref{tab:attestation_threats} summarizes the major security management threats associated with Secure Boot and Remote Attestation in AI-native 6G networks.

\section{Countermeasures and Open Challenges}The cross-layer threat taxonomy presented in the previous section demonstrates that security and privacy risks in AI-native \ac{6G} networks extend beyond individual technologies and propagate across interconnected layers. Addressing these threats therefore requires a unified set of countermeasures that provide coordinated protection throughout the network life cycle. Guided by the proposed unified security and privacy framework, this section discusses mapping of layered threats to countermeasures, the key cross-layer security functions—including identity and access management, security orchestration and automation, AI governance and assurance, privacy governance and compliance, and standards harmonization—and outlines the open challenges that remain toward achieving secure, trustworthy, and interoperable AI-native \ac{6G} ecosystems.

\subsection{Mapping of Layered Threats to Countermeasures in AI-Native 6G Networks}
The diverse threat landscape of AI-native 6G networks necessitates a multi-layered and coordinated defense strategy. As shown in Table~\ref{tab:threat_countermeasure_mapping}, threats arising across infrastructure, network and architectural, AI, privacy, and security management domains require corresponding security and privacy mechanisms tailored to their unique characteristics. While infrastructure threats can be mitigated through physical-layer security and secure communication techniques, network and architectural threats require robust identity management, access control, and security orchestration mechanisms. Similarly, AI-related threats demand governance and assurance frameworks to ensure the robustness, transparency, and trustworthiness of intelligent systems. Privacy threats necessitate privacy-enhancing technologies and regulatory compliance mechanisms, whereas security management threats require advanced cryptographic protections, platform integrity verification, and continuous authentication. Collectively, these countermeasures provide the foundation for the proposed unified security and privacy framework, enabling end-to-end protection across heterogeneous AI-native 6G environments.

\begin{table*}[t]
\centering
\caption{Mapping of Layered Threats to Countermeasures in AI-Native 6G Networks}
\label{tab:threat_countermeasure_mapping}
\renewcommand{\arraystretch}{1.15}
\begin{tabular}{|p{3cm}|p{4cm}|p{7cm}|}
\hline
\textbf{Threat Layer} &
\textbf{Representative Threats} &
\textbf{Key Countermeasures} \\
\hline

Infrastructure Threats &
RIS attacks, JCAS information leakage, NTN spoofing and jamming, XL-MIMO eavesdropping &
Physical Layer Security (PLS), secure beamforming, secure sensing, authentication mechanisms, spectrum protection, anti-jamming techniques \\
\hline

Network \& Architectural Threats &
O-RAN interface attacks, network slice breaches, MEC compromise, intent manipulation &
Identity and Access Management, Intelligent ZTA, secure APIs, network isolation, remote attestation, secure orchestration and automation \\
\hline

AI Threats &
Data poisoning, adversarial attacks, model extraction, agent manipulation, model inversion &
AI Governance and Assurance, secure model life cycle management, adversarial training, explainable AI, federated learning security, continuous model monitoring \\
\hline

Privacy Threats &
Location tracking, inference attacks, digital twin leakage, behavioral profiling, metadata leakage &
Differential Privacy, Federated Learning, Homomorphic Encryption, secure multiparty computation, privacy governance and compliance, data minimization \\
\hline

Security Management Threats &
Identity spoofing, credential compromise, cryptographic attacks, platform integrity violations &
Identity and Access Management, Post-Quantum Cryptography, Secure Boot, Remote Attestation, continuous verification, credential management \\
\hline

\end{tabular}
\end{table*}
Table~\ref{tab:threat_countermeasure_mapping} illustrates how the diverse threats identified across the proposed cross-layer taxonomy can be mitigated through a set of complementary security and privacy mechanisms. While individual countermeasures address specific vulnerabilities within particular domains, effective protection of AI-native 6G networks requires coordinated security functions that operate across multiple layers. Consequently, the following subsections discuss four key cross-layer security functions—Identity and Access Management, Security Orchestration and Automation, AI Governance and Assurance, and Privacy Governance and Compliance—which collectively form the foundation of the proposed unified security and privacy framework.

\subsection{Cross-Layer Countermeasure-Identity and Access Management}
Identity and Access Management (IAM) serves as a foundational countermeasure for mitigating cross-layer security threats in AI-native 6G networks by enabling continuous authentication, authorization, and access control across heterogeneous users, devices, AI agents, and network functions. As 6G ecosystems become increasingly decentralized and autonomous, robust IAM mechanisms are essential for preventing identity spoofing, unauthorized access, privilege escalation, and cross-domain security breaches. Emerging approaches such as iZTA, decentralized identity management, multi-factor authentication (MFA), and continuous verification provide adaptive and context-aware access control capabilities suitable for dynamic 6G environments \cite{chen2023zero,nahar2024survey,garzon2022decentralized}. Furthermore, integrating IAM with remote attestation, credential management, and AI-assisted risk assessment can strengthen trust establishment and secure interactions across infrastructure, network, AI, privacy, and security management domains \cite{sedjelmaci2023enabling,wei2023zero}.

Despite significant advances in identity and access management, several challenges remain for AI-native 6G networks. First, the massive scale and heterogeneity of 6G ecosystems, encompassing terrestrial, aerial, satellite, edge, and AI-driven domains, make unified identity management and interoperability across administrative boundaries a difficult task. Second, continuous authentication and context-aware access control may introduce substantial signaling and computational overhead, particularly for resource-constrained devices and ultra-low-latency applications. Third, the increasing adoption of autonomous AI agents, digital twins, and multi-agent systems necessitates new identity frameworks capable of securely managing machine-to-machine interactions while preserving privacy and accountability. Addressing these challenges will be critical for realizing scalable, interoperable, and trustworthy identity and access management in AI-native 6G networks.

\subsection{Cross-Layer Countermeasure-Security Orchestration and Automation}
Security Orchestration and Automation (SOA) is a key enabler for managing the complexity of AI-native 6G networks, where security decisions must be coordinated across heterogeneous infrastructures, network domains, AI systems, and privacy-sensitive applications. By integrating automated threat detection, policy enforcement, incident response, and life cycle management, SOA enables real-time adaptation to evolving cyber threats while reducing operational overhead. Emerging frameworks such as Zero-Touch Service Management (ZSM), intent-driven security management, and AI-assisted orchestration facilitate autonomous security operations through continuous monitoring and dynamic policy updates \cite{ETSI_ISG_ZSM_AR_2023,nguyen2024emerging}. Furthermore, security orchestration can enhance cross-domain visibility and enable coordinated responses to attacks spanning infrastructure, network, AI, and privacy layers, thereby improving the resilience and adaptability of future 6G ecosystems \cite{porambage2024security,chen2023zero}.

Despite its potential, several challenges hinder the realization of effective security orchestration and automation in AI-native 6G networks. Firstly, achieving interoperability among heterogeneous technologies, vendors, and administrative domains remains difficult due to differing security policies, interfaces, and operational requirements. Secondly, the increasing reliance on AI-driven orchestration introduces risks associated with adversarial manipulation, explainability, and accountability of automated security decisions. Thirdly, ensuring real-time threat detection and coordinated response across highly distributed environments, including edge, cloud, terrestrial, and non-terrestrial networks, presents significant scalability challenges. Finally, balancing automation with human oversight remains an open issue, particularly in safety-critical scenarios where incorrect security decisions may have widespread operational consequences.

\subsection{Cross-Layer Countermeasure-AI Governance and Assurance}
AI Governance and Assurance play a critical role in mitigating security and privacy threats arising from the widespread integration of artificial intelligence into AI-native 6G networks. As AI technologies such as federated learning, foundation models, AI-RAN, multi-agent systems, and digital twins become integral to network operations, ensuring their trustworthiness and robustness is essential. AI governance establishes policies, accountability mechanisms, and regulatory controls for the development, deployment, and operation of AI systems, while AI assurance focuses on validating their security, reliability, explainability, and resilience against adversarial manipulation. Techniques such as explainable AI (XAI), model auditing, adversarial robustness testing, secure model life cycle management, and continuous monitoring can help mitigate risks including data poisoning, model evasion, privacy leakage, and autonomous decision manipulation \cite{wang2026ai, ai2023artificial, act2024eu, goodfellow2015explainingharnessingadversarialexamples}. By providing transparency, accountability, and risk-aware AI management, AI governance and assurance strengthen the security and trustworthiness of AI-enabled 6G ecosystems.

Despite growing interest in trustworthy AI, several challenges remain for AI governance and assurance in AI-native 6G networks. Firstly, the distributed and autonomous nature of AI systems operating across edge, cloud, terrestrial, and non-terrestrial domains complicates monitoring, auditing, and accountability. Secondly, ensuring robustness against emerging adversarial attacks on foundation models, federated learning systems, and multi-agent environments remains an active research challenge. Thirdly, achieving explainability and transparency for large-scale AI models while maintaining operational efficiency is difficult, particularly in real-time network management scenarios. Finally, the absence of harmonized standards and governance frameworks for AI security, privacy, and accountability creates challenges for interoperability, regulatory compliance, and cross-domain trust in future 6G ecosystems.
\subsection{Cross-Layer Countermeasure-Privacy Governance and Compliance}
Privacy Governance and Compliance are essential for addressing the growing privacy risks introduced by AI-native 6G networks, where massive volumes of personal, contextual, sensing, and operational data are continuously collected, processed, and shared across distributed environments. Emerging technologies such as JCAS, digital twins, edge intelligence, foundation models, and immersive XR applications significantly increase the risk of unauthorized data collection, inference attacks, identity linkage, and behavioral profiling. Privacy governance establishes policies, accountability mechanisms, and data management practices that ensure the lawful and ethical use of data, while compliance frameworks provide adherence to regulatory requirements such as GDPR, AI governance regulations, and sector-specific privacy standards \cite{GDPR2019, dass2024addressing, qu2024privacy}. Furthermore, privacy-enhancing technologies, including differential privacy, federated learning, homomorphic encryption, secure multiparty computation, and data minimization techniques, can help mitigate privacy risks while supporting intelligent 6G services \cite{dwork2014algorithmic,wang2026aisecurity}. By integrating technical, organizational, and regulatory safeguards, privacy governance and compliance contribute to building trustworthy and privacy-preserving AI-native 6G ecosystems.

Despite significant advances in privacy-preserving technologies and regulatory frameworks, several challenges remain for privacy governance in AI-native 6G networks. Firstly, the convergence of communication, sensing, computing, and AI creates complex data ecosystems where personal and contextual information may be collected and inferred without explicit user awareness. Secondly, balancing privacy protection with the need for real-time intelligence, personalization, and autonomous decision making remains a major challenge, particularly in latency-sensitive applications. Thirdly, ensuring consistent privacy policies and regulatory compliance across multiple jurisdictions, operators, and service providers is difficult in globally interconnected 6G environments. Finally, emerging technologies such as digital twins, foundation models, and multi-agent systems introduce new forms of privacy leakage and inference risks that are not yet fully addressed by existing governance frameworks and standards.
\subsection{Cross-Layer Countermeasure-Standardization Harmonization Analysis}\label{standards}\begin{table*}[t]
\centering
\caption{Standards Harmonization Matrix for Security and Privacy in AI-Native 6G Networks}
\label{tab:harmonization}
\resizebox{\textwidth}{!}{
\begin{tabular}{|p{3.4cm}|c|c|c|c|c|c|c|p{4.8cm}|}
\hline
\textbf{Security/Privacy Function}
& \textbf{3GPP}
& \textbf{ETSI}
& \textbf{ITU}
& \textbf{NIST}
& \textbf{IETF}
& \textbf{O-RAN}
& \textbf{AI-RAN}
& \textbf{Gap Analysis} \\
\hline

Identity Management
& \checkmark
& P
& P
& --
& P
& P
& --
& Lack of unified cross-domain identity framework for terrestrial, NTN, and AI-native environments. \\
\hline

Zero Trust Architecture
& P
& \checkmark
& --
& \checkmark
& --
& \checkmark
& --
& No harmonized implementation model across 6G ecosystems. \\
\hline

AI Security \& Assurance
& P
& \checkmark\checkmark
& --
& P
& --
& \checkmark
& \checkmark
& Absence of standardized AI assurance and model trustworthiness framework. \\
\hline

AI Agent Security
& --
& \checkmark
& --
& --
& --
& P
& \checkmark
& Significant standardization gap for autonomous AI agents. \\
\hline

Privacy Protection
& P
& \checkmark\checkmark
& \checkmark
& --
& \checkmark
& P
& --
& Lack of unified privacy governance and consent management mechanisms. \\
\hline

JCAS Security
& P
& \checkmark\checkmark
& P
& --
& --
& --
& --
& Fragmented approaches to sensing security and privacy. \\
\hline

NTN Security
& P
& P
& \checkmark
& --
& --
& --
& --
& No common trust architecture across terrestrial and non-terrestrial domains. \\
\hline

Network Slicing Security
& \checkmark\checkmark
& \checkmark
& --
& --
& --
& P
& \checkmark
& Limited interoperability guidance for cross-domain slice security. \\
\hline

O-RAN Security
& P
& P
& --
& --
& --
& \checkmark\checkmark
& P
& O-RAN controls are not fully aligned with 3GPP security models. \\
\hline

Post-Quantum Security
& P
& \checkmark
& \checkmark
& \checkmark\checkmark
& P
& P
& --
& Lack of migration and interoperability roadmap. \\
\hline

Physical Layer Security
& P
& \checkmark
& P
& --
& --
& --
& --
& Not systematically integrated into standards frameworks. \\
\hline

\end{tabular}
}
\vspace{1mm}
\footnotesize{\textbf{Legend:}
\checkmark\checkmark = Strong focus,
\checkmark = Explicit coverage,
P = Partial coverage,
-- = Limited or no explicit coverage.}
\end{table*}

\begin{table*}[t]
\centering
\caption{Proposed Harmonization Priorities for AI-Native 6G Security and Privacy}
\label{tab:priorities}
\resizebox{\textwidth}{!}{
\begin{tabular}{|p{2.8cm}|p{4.5cm}|p{4.5cm}|p{5.5cm}|}
\hline

\textbf{Domain}
& \textbf{Current State}
& \textbf{Key Challenge}
& \textbf{Proposed Harmonization Direction} \\
\hline

AI Security
& ETSI SAI, O-RAN, and AI-RAN address AI security independently.
& Inconsistent AI trust, assurance, and risk assessment mechanisms.
& Develop a unified AI assurance and certification framework for AI-native 6G. \\
\hline

AI Agents
& Early work emerging in ETSI ENI and AI-RAN.
& No security baseline for autonomous AI agents.
& Establish standards for agent authentication, trust, explainability, and accountability. \\
\hline

NTN Security
& Security efforts split between ITU and 3GPP.
& Lack of common trust model across space-air-ground networks.
& Develop unified NTN trust, authentication, and key-management framework. \\
\hline

JCAS Security
& ETSI and 3GPP have initiated JCAS-related activities.
& Privacy, consent, and sensing data protection remain fragmented.
& Define common sensing privacy and authorization framework. \\
\hline

Identity Management
& Multiple identity approaches exist across standards.
& Fragmented credentials and trust anchors.
& Introduce federated identity and decentralized trust architecture. \\
\hline

Network Slicing
& Security mechanisms defined independently by multiple organizations.
& Cross-domain slice security and trust interoperability.
& Standardize end-to-end slice trust and isolation policies. \\
\hline

Post-Quantum Security
& NIST, ETSI, ITU, and 3GPP actively pursuing PQC migration.
& Lack of coordinated deployment roadmap.
& Develop interoperable migration and crypto-agility guidelines. \\
\hline

Privacy Governance
& Privacy mechanisms differ across organizations and use cases.
& No common privacy model for AI-native and sensing-enabled networks.
& Harmonize privacy-preserving architectures, consent management, and data governance. \\
\hline

\end{tabular}
}
\end{table*}

The realization of secure and trustworthy AI-native \ac{6G} networks requires coordinated standardization across multiple organizations, including \ac{3GPP}, \ac{ETSI}, \ac{ITU}, \ac{NIST}, \ac{IETF}, the O-RAN Alliance, and the AI-RAN Alliance. These organizations collectively address security for communication infrastructures, cloud-native architectures, AI systems, privacy-preserving technologies, and post-quantum cryptography. 

Beyond formal standardization bodies, several policy and strategic initiatives are shaping the security and privacy landscape of AI-native 6G. Recent examples include Global Coalition on Telecommunications (GCOT) \cite{GCOT_techradar} Security and Resilience Principles for 6G, which advocate security-by-design, trustworthy AI, supply-chain security, interoperability, and resilience as foundational requirements for future networks. Similarly, the \ac{ITU} \ac{IMT}-2030 Framework, the \ac{NIST} AI security and post-quantum cryptography initiatives, the \ac{NGMN} 6G architecture vision, and the European \ac{6G-IA} vision emphasize secure, trustworthy, and interoperable AI-native communication systems. Collectively, these initiatives complement technical standards by providing policy guidance and strategic direction for the evolution of secure 6G ecosystems.

Table~\ref{tab:harmonization} summarizes the current coverage of key security and privacy functions across major standardization bodies, while Table~\ref{tab:priorities} identifies the corresponding harmonization priorities for AI-native 6G networks. Although significant progress has been made, important gaps remain in AI assurance, AI-agent security, \ac{JCAS} privacy, \ac{NTN} trust, cross-domain identity management, and post-quantum migration. Moreover, many security mechanisms continue to evolve independently, increasing the risk of inconsistent trust models, privacy frameworks, and interoperability challenges. Future standardization efforts should therefore prioritize unified approaches for AI governance, federated identity management, privacy-preserving sensing, crypto-agility, and end-to-end security across heterogeneous terrestrial and non-terrestrial environments, while aligning with emerging policy initiatives.

\section{Future Research Directions}
\label{sec:future}

The unified security and privacy framework presented in this survey establishes a holistic foundation for addressing cross-layer security challenges in AI-native \ac{6G} networks. However, realizing secure, trustworthy, and globally interoperable 6G ecosystems requires continued advances beyond the proposed framework. Several promising research directions are outlined below.

\begin{itemize}
\item{\bf{Framework Validation and Intelligent Security Orchestration:}}
Future work should focus on validating the proposed framework through large-scale testbeds, digital twins, and AI-native network platforms. In addition to experimental evaluation, intelligent security orchestration capable of correlating threats, countermeasures, and security policies across infrastructure, network, AI, privacy, and management layers will be essential for practical deployment.
\item{\bf{Trustworthy and Autonomous Security:}}
Future AI-native networks should evolve toward self-protecting and self-healing security architectures capable of continuous risk assessment, adaptive policy enforcement, and autonomous incident response. Achieving this vision requires trustworthy AI, explainable decision making, secure multi-agent collaboration, and human oversight to ensure transparency and accountability.
\item{\bf{Global Standards Convergence and Security Certification:}}
Future research should bridge the gap between technical solutions and global deployment by translating harmonized security principles into interoperable standards, certification frameworks, and common trust models. Strengthening collaboration among standards organizations, industry alliances, and emerging policy initiatives will be critical to ensuring secure, interoperable, and resilient AI-native \ac{6G} ecosystems.
\item{\bf{Security Benchmarks and Quantitative Evaluation:}}
The lack of standardized security benchmarks limits objective comparison of emerging protection mechanisms for AI-native \ac{6G}. Future research should establish common datasets, attack models, evaluation metrics, and reproducible benchmarking platforms to assess the effectiveness, scalability, and interoperability of cross-layer security solutions.

\end{itemize}
\section{Conclusion}
AI-native \ac{6G} networks will enable unprecedented levels of intelligence, automation, and connectivity by integrating advanced communication, computing, sensing, and AI technologies. However, this convergence also introduces a highly complex and evolving security and privacy landscape, characterized by threats spanning infrastructure, network architectures, AI systems, privacy-sensitive applications, and security management mechanisms.

In this paper, we presented a comprehensive survey of security and privacy challenges in AI-native \ac{6G} networks. We first highlighted the fragmentation of security and privacy requirements across emerging technologies, architectures, AI frameworks, and standards. To address this challenge, we proposed a unified security and privacy framework and developed a cross-layer threat taxonomy covering infrastructure, network and architectural, AI, privacy, and security management domains. Furthermore, we mapped key threats to corresponding countermeasures and discussed the role of standards harmonization in enabling interoperable and trustworthy 6G ecosystems.

Our analysis demonstrates that securing AI-native \ac{6G} networks requires a holistic and cross-layer approach rather than isolated technology-specific solutions. Future advancements in trustworthy AI, autonomous security orchestration, privacy-preserving intelligence, quantum-resilient cryptography, and standards harmonization will be critical to realizing secure, resilient, and trustworthy next-generation communication networks.
\section*{Acknowledgment}
This work is supported by EPSRC and DSIT funded project - CHEDDAR: Communications Hub For Empowering Distributed Cloud Computing Applications And Research (EP\-/X040518/1), (EP/Y037421/1) and EPSRC funded project REMOTE (EP/Y019229/1).  All content was critically reviewed and verified by the authors.

%\bibliographystyle{ieeetr}
%\bibliography{references_security}
%\balance
\printbibliography
%\appendix
\end{document}